\documentclass[fleqn,usenatbib]{mnras}

\usepackage{newtxtext,newtxmath}

\usepackage[T1]{fontenc}

\DeclareRobustCommand{\VAN}[3]{#2}
\let\VANthebibliography\thebibliography
\def\thebibliography{\DeclareRobustCommand{\VAN}[3]{##3}\VANthebibliography}

\usepackage{comment}
\usepackage{hyperref}
\usepackage{xcolor}
\usepackage{graphicx}	
\usepackage{hhline}
\usepackage{amsmath}	
\usepackage{multirow}
\usepackage{silence}
\usepackage{subcaption} 
\PassOptionsToPackage{pdfpagelabels=false}{hyperref} 

\title[Cosmic Web Quenching of Central Galaxies]{The Imprint of Cosmic Web Quenching on Central Galaxies}

\author[N. Winkel et al.]{
N. Winkel,$^{1,2}$\thanks{E-mail: winkel@mpia.de}
A. Pasquali,$^{2}$
K. Kraljic,$^{3,4}$
R. Smith,$^{5}$
A. Gallazzi,$^{6}$
T. M. Jackson$^{2}$
\\
$^{1}$Max Planck Institute for Astronomy, K\"onigstuhl 17, D-69117 Heidelberg, Germany\\
$^{2}$Astronomisches Rechen-Institut, Zentrum f\"ur Astronomie der Universit\"at Heidelberg, M\"onchhofstr. 12 - 14, D-69120 Heidelberg, Germany\\
$^{3}$Institute for Astronomy, Royal Observatory, Edinburgh EH9 3HJ, United Kingdom\\
$^{4}$Aix Marseille Universit\'e, CNRS, CNES, Laboratoire d'Astrophysique de Marseille, Marseille, France\\
$^{5}$Korea Astronomy and Space Science Institute, 776 Daedeokdae-ro, Yuseong-gu, Daejeon 34055, Korea\\
$^{6}$INAF-Osservatorio Astrofisico di Arcetri, Largo Enrico Fermi 5, I-50125 Firenze, Italy\\
}

\date{Accepted 2021 May 25. Received 2021 May 03; in original form 2021 March 03}

\pubyear{2020}

\begin{document}
\label{firstpage}
\pagerange{\pageref{firstpage}--\pageref{lastpage}}
\maketitle

\begin{abstract} 
    We investigate how cosmic web environment impacts the average properties of central galaxies in the Sloan Digital Sky Survey (SDSS). We analyse how the average specific star-formation rate, stellar age, metallicity and element abundance ratio [$\alpha$/Fe] of SDSS central galaxies depend on distance from the cosmic web nodes, walls and filaments identified by DisPerSE.
    In our approach we control for galaxy stellar mass and local density differentiated between field and group environment. Our results confirm the known trend whereby galaxies exhibit lower specific star-formation rates with decreasing distance to the cosmic web features. Furthermore, we show that centrals closer to either nodes, walls or filaments are on average older, metal richer and $\alpha$-enhanced compared to their equal mass counterparts at larger distances. The identified property gradients appear to have the same amplitude for central galaxies in the field as for those in groups. 
    Our findings support a cosmic web quenching that stems from {\it nurture} effects, such as ram pressure stripping and strangulation, and/or {\it nature} effects linked to the intrinsic properties of the cosmic web.
\end{abstract}




\section{Introduction}
\label{stn:Introduction}
In the paradigm of the $\Lambda$CDM cosmology, structures arise from the primordial density fluctuations that were imprinted in the dark matter distribution after the cosmic inflation. 
The power spectrum as predicted by the cosmological model dictates a structure growth in a hierarchical manner. As a consequence of the anisotropic collapse of the density perturbations, galaxies in the present day universe are scattered in an intricately connected web-like pattern \citep{Zel'dovich1970}, known as the `cosmic web' \citep{Bond1996}. A major challenge for the galaxy formation and evolution theory is explaining the broad range of galaxy types, which is a reflection of the numerous physical processes involved in their formation and evolution. 

Although internal processes, as characterised by galaxy stellar mass $M_\star$, appear to be the main drivers of galaxy evolution \citep{Gavazzi1996, Mouhcine2007, Pasquali2010,Peng2010, Hughes2013, Gallazzi2021}, it is well-established that the environment in which galaxies reside also plays an important role in their evolution. Already four decades ago, \cite{Dressler1980} showed that galaxy morphology is a continuous function of the projected local galaxy density. Several following studies have identified correlations of colour, star formation rate, stellar ages and metallicities with environmental density \citep[e.g.][]{Hashimoto1998, Lewis2002, Blanton2003, Weinmann2006, Burton2013}. \cite{Peng2010} find that that low-mass galaxies ($\textrm{log}(M_\star / (h^{-1}\textrm{M}_\odot)) < 10.6$) in particular are sensitive to the density wherein they reside.

The advent of group finder algorithms \citep[e.g.][]{Yang2005}
allowed to classify galaxies in large sky surveys into centrals and less massive satellites which orbit in the group/cluster potential. In this framework central galaxies play a special role as they are the dominating galaxy in their environment.
It has been found that at fixed stellar mass $M_\star$ the fraction of actively star forming centrals (with specific star formation rate $\textrm{log}(\textrm{sSFR}/\textrm{yr}^{-1}) > -11$) is higher than for the satellites \citep{vandenBosch2008, Pasquali2009, Wetzel2012}. \cite{Pasquali2010} have shown that at fixed stellar mass the median stellar ages and metallicities of central galaxies increase with increasing halo mass. In massive haloes, central galaxies are observed to be more extended than satellites with the same stellar mass \citep{Bernardi2009, Vulcani2014} which can be interpreted as the result of AGN feedback in massive central galaxies, whereby gas outflows dynamically heat the inter-stellar medium (ISM). This eventually leads to an adiabatic expansion of the stellar component \citep{Choi2018}. A complementary hypothesis suggests that centrals accumulate stars in their outskirts by a series of minor mergers which enable them to grow in size and mass \citep{Naab2009, Oser2010, Huang2018, Jackson2021}. The radial gradients observed in the stellar populations of massive early-type centrals are in agreement with this scenario, as the average stellar metallicity decreases and stars tend to be older with increasing galactocentric distance \citep{Greene2013, LaBarbera2017}.
Furthermore, the most massive group member central galaxies can accrete gas and stars that have been stripped of their satellite galaxies, enabling them to grow in both size and mass over a long time.

While many studies focus on particularly dense environments such as clusters, transformation processes may already occur in less dense environments. It has been noticed that strangulation and ram-pressure stripping acting on group/cluster satellite galaxies cannot ultimately explain the occurrence of galaxies with a suppressed SFR at large distances from cluster centres \citep{Kodama2001, Lewis2002, Sarron2019}. Several works suggest that galaxies experience environmental erosion effects before being accreted onto a cluster, for instance, while living in their previous host group \citep[e.g.][]{Smith2012, DeLucia2012, RobertsParker2017, Pasquali2019}. In the literature this phenomenon of \emph{pre-processing} is often used to explain galaxies having their SF quenched well before crossing any cluster virial radius.
This hypothesis has been supported early on by an analytical model from \citet{Fujita2004} suggesting that during hierarchical clustering, late-type galaxies have experienced environmental erosion effects in small groups prior to entering the cluster virial radius. Numerical simulations have allowed to study this phenomenon in more detail and come to the conclusion that the previous group host halo can affect a significant fraction of present day cluster satellites before their accretion \citep{Taranu2014, Han2018}. \citet{DeLucia2012} find the highest fraction of pre-processed galaxies amongst low-mass galaxies, of which 28\%  have been accreted from haloes more massive than $10^{13}\,h^{-1}\textrm{M}_\odot$. The quenching efficiency in mildly dense environments remains however unclear.

The cosmic web itself may be a source of pre-processing. Hydrodynamical simulations suggest that up to half of the universe baryon content could be in the form of shock-heated gas between clusters of galaxies \citep{CenOstriker2006}. The cosmic web filaments are observed to hold a substantial amount of this so-called Warm Hot Intergalactic Medium with temperatures varying between 10$^5$ and 10$^7$K \citep{Clampitt2016, Tanimura2019, Tanimura2020}. This component of the cosmic web may lead to ram-pressure stripping that acts on the traversing galaxies, an effect referred to as \emph{cosmic web stripping}. Although ram-pressure stripping might enhance the star formation temporarily, the removal of cold gas effectively reduces the galaxy SFR. This scenario is further complicated by the vast gas reservoirs contained in the filaments. Continuous gas streams provide a large gas supply that potentially fuels galaxies and continuously enhances their SFR \citep{Darvish2014}. Although the density gradients within the cosmic web are shallower than within groups and clusters, these processes are expected to be imprinted in the galaxy properties to some degree.
\par
Observations \citep{Tempel2013,  TempelLibeskind2013, Zhang2013, Kraljic2020b} and simulations \citep{Pichon2011, Codis2012, Laigle2015} have shown that galaxy kinematics depend on the position in the large scale structure. While recently formed low-mass galaxies have their spins preferentially aligned with their neighbouring filaments, the spins of more evolved galaxies in the vicinity of filaments are typically perpendicular to the filament axis.
Since the galaxy number density in filaments is higher, galaxy interactions such as mergers occur more frequently in filaments than in the field \citep{Lhuillier2012}. 
From the kinematics of the tidal torque field a picture emerges where low-mass haloes grow during their accretion onto filaments and align their spin parallel due to filamentary flows that advect angular momentum. The next generation of galaxies, with spins preferentially perpendicular to the filaments, are formed by successive mergers along the filament as they migrate towards the nodes of the cosmic web. \citep[e.g.][]{Pichon2011, Aragon-Calvo2014, Dubois2014, Welker2014, Kang2015, Codis2015, Wang2018}. In this framework, the morphological complexity of the present day cosmic web is a reflection of the non-linearities in the formation and evolution of angular momentum in the dark matter haloes. 
\par\noindent
By applying the {\small DisPerSE} cosmic web identification method to the GAMA spectroscopic survey \citep{Driver2009}, \citet{Kraljic2018} analysed the dependence of the dust corrected $u-r$ colours and sSFR as a function of galaxies 3D distances to the cosmic web features (i.e. nodes, walls and filaments). Out of the whole population of galaxies, they find passive galaxies to be significantly closer to the cosmic web filaments and walls than their star-forming counterparts. Furthermore, they report a mass segregation among the star-forming population alone where massive galaxies have on average smaller distances to each of the features. At fixed stellar mass the colours redden and the sSFR decreases with decreasing distance to both filaments and walls which leads the authors to conclude that quenching processes must be at work.\\
We build upon this work, by analysing how galaxy physical properties such as average stellar age, metallicity and element abundance ratio [$\alpha$/Fe] of central galaxies vary as a function of their position in the cosmic web. We focus on central galaxies, either in the field or in group/cluster environments, since they are less affected by the environmental processes due to their host haloes, and thus allow us to better single out the effects of the cosmic web on their properties.\\
In Section~\ref{stn:Data} we present the galaxy data in detail, while in Section~\ref{stn:Methodology} we describe the methodology adopted to analyse galaxy
properties as a function of distance to the cosmic web's nodes, walls and filaments. We present and discuss our results
in Section~\ref{stn:Results} and \ref{stn:Discussion}, respectively, and draw our summary in Section~\ref{stn:Summary}.
\par\noindent
Throughout this work we adopt a flat WMAP3 $\Lambda$CDM cosmology \citep{Spergel2007} with the Hubble constant $H_0 = 67.5\,\textrm{km}\textrm{s}^{-1}\,\textrm{Mpc}^{-1}$, the parameters total matter density $\Omega_m = 0.238$, baryon density $\Omega_b=0.042$, dark energy density $\Omega_{\Lambda} = 0.762$ and $\sigma_8 = 0.75$ and $n=0.951$. Luminosity and distances are scaled with $h$. We refer to the 10-based logarithm by using $\textrm{log}$.

\section{Data}
\label{stn:Data}

\subsection{Group Catalogues}
We use the halo mass given in the group catalogue from \citet{Wang2014} to quantify the local environment of galaxy groups and clusters. This catalogue is based on the adaptive halo-based group finder algorithm by \cite[][hereafter Y07]{Yang2007}, updated for SDSS data release 7 (DR7, \citealt{Abazajian2009}). Galaxy stellar masses $M_\star$ were computed by Y07 using the relation between stellar-mass-to-light ratio and colour from \cite{Bell2003}.\\
The Y07 algorithm uses the friends-of-friends algorithm \citep{Davis1985} to identify geometrical, mass-weighted centres of tentative groups. Group membership is then established using an iterative approach that re-assigns group membership based on the virial radius spanned by the halo mass. The corresponding halo masses are estimated using the ranking of the halo's characteristic stellar mass, defined as the total stellar mass of all group members with \mbox{$^{0.1}M_r - 5 \textrm{log}h \leq -19.5\,\textrm{mag}$}. With this method, halo masses can only be estimated for groups more massive than \mbox{$10^{12} \, h^{-1}\textrm{M}_\odot$} (corresponding to at least one member with \mbox{$^{0.1} M_r - 5 \textrm{log} h \leq -19.5\,\textrm{mag}$}). For less massive systems down to \mbox{$M_h \approx 10^{11}h^{-1}\textrm{M}_\odot$}, halo masses are estimated using an extrapolation of the relation between stellar mass of central galaxies and the halo mass of their groups.
Y07 report that the average scatter of the group masses decreases from $0.1\,\textrm{dex}$ at the low mass end to $0.05\,\textrm{dex}$ at the massive end. However, \citet{Yang2005} and \citet{Weinmann2006} used mock galaxy redshift surveys to show that the group finder algorithm later employed by Y07 and \citet{Wang2014} returns accurate average halo properties.\\
Applying the group finder algorithm to SDSS DR7 NYU-VAGC, \citet{Wang2014} obtained 472,416 groups ranging from isolated galaxies to rich galaxy clusters. 
Galaxies which are the most massive members of their group are referred to as \emph{central} galaxies, while all the other group members are referred to as \emph{satellite} galaxies.
However, the majority of the groups only contain one member galaxy and only 23,700 have more than two member galaxies.

\subsection{Galaxy Properties}
\label{stn:Galaxy Properties}

\subsubsection{Star Formation Rates}
\label{stn:Star Formation Rates} 
We retrieve the global specific star formation rates \mbox{$\textrm{sSFR}_\textrm{glo} = \textrm{SFR}_{\textrm{glo}} / M_\star$} from \citet{Brinchmann2004} which were computed by fitting the SDSS photometry of the outer regions of galaxies with simple stellar population (SSP) models published by \citet[][BC03]{BruzualCharlot2003} convolved with the extinction law from \citet{CharlotFall2000}. 
Since galaxies may have a significant fraction of star formation on-going beyond the radius covered by the fibre, $\textrm{sSFR}_\textrm{glo}$ is expected to give a sensible estimate of the galaxies' total star formation activity. 

\subsubsection{Stellar Population Properties}
\label{stn:Stellar Populations}
We use the catalogue of stellar ages, metallicities and element abundance ratios [$\alpha$/Fe] from \citet{Gallazzi2021} which were computed following the same procedure as described in \citet{Gallazzi2005, Gallazzi2006}.
The stellar population parameters are constrained by the spectral indices $\textrm{D4000}_\textrm{n}$, $\textrm{H}\beta$, $\textrm{H} \delta_A$ + H$\gamma _A$, [MgFe]' and [$\textrm{Mg}_2$Fe] which were measured in the 3 arcsec fibres of the SDSS spectrograph. The probability density functions (PDFs) of the r-band luminosity-weighted age and metallicity were estimated by comparing the observed indices with those of model spectra from BC03 convolved with a library of star formation histories (SFHs) and metallicities.\par\noindent

The [$\alpha$/Fe] estimate is based on the excess in the observed index ratio Mgb/<Fe> with respect to that of the BC03-based models. The calibration of the excess in $\textrm{Mgb} / \langle \textrm{Fe} \rangle$ with [$\alpha$/Fe] is shown to be largely independent of age and metallicity in particular for age $> 1\,\textrm{Gyr}$ and $Z_\star > 0.5\,\textrm{Z}_\odot$ and of the models used for the calibration \citep[see][for more details]{Gallazzi2021}.\par\noindent
The methodology by which the [$\alpha$/Fe] abundances were computed is described in \citet{Gallazzi2006}. Here, the ratio is quantified by the Mg and Fe absorption line indices which are sensitive to [$\alpha$/Fe] \citep{Thomas2003}. The Mgb/$\langle \textrm{Fe} \rangle$ ratio was compared to the model that best reproduces the five spectral indices used to compute age and metallicity (see above), where $\langle \textrm{Fe} \rangle$ is the average between the Fe5270 and Fe5335 index strengths. Since the abundance ratios of the models are solar-scaled, any deviation $\Delta (\textrm{Mgb} / \langle \textrm{Fe} \rangle)$ between observed index strength and model index strength is interpreted as an excess of $\alpha$-elements with respect to the solar value. For each model the difference $\Delta (\textrm{Mgb} / \langle \textrm{Fe} \rangle)$ was computed at fixed age and metallicity. With the SSPs from \citet{Thomas2003}, $\Delta (\textrm{Mgb} / \langle \textrm{Fe} \rangle)$ was then calibrated against [$\alpha$/Fe] as a function of age and metallicity.
For each parameter we take the median of the PDF as fiducial value and half of the 16-84 percentile range as the uncertainty.

\subsection{The Cosmic Web Metric}
\label{stn:The Cosmic Web Metric}
We adopt the galaxies' distances to the cosmic web features computed as in \cite{Kraljic2020b} using the Discrete Persistent Structure Extractor ({\small DisPerSE}, \citealt{Sousbie2011}) to extract nodes, filaments and walls from the galaxy distribution of SDSS DR7.
Due to the Doppler shift caused by the random peculiar velocities of galaxies within groups and clusters, the galaxy distribution is elongated in redshift space. \cite{Kraljic2020b} used the \cite{Tempel2014} SDSS catalogue which was corrected for this so-called Finger-of-God effect.\\ 
{\small DisPerSE} allows a scale and parameter-free identification of 3D features of a galaxy spatial distribution. Using Morse theory and persistence theory\footnote{Topological persistence is a method to quantify the importance of topological features when noise is present. In this way the topology can be categorised by its robustness.}, it exploits the properties of the topology that is induced by a discrete set of data points. A detailed description of the algorithm can be found in \citet{Sousbie2011} and illustrations in an astrophysical context in \citet{SousbiePichonKawahara2011}. \\
Figuratively, voids are surrounded by sheet-like walls, which again are framed by filaments which bridge from one high density knot to another. By construction, walls are the most frequently occurring cosmic web features occupying the largest share of space followed by filaments and nodes. \\
Given its topological approach, {\small DisPerSE} does not consider the sampling of the data set. Instead, the significance of the extracted structures is quantified by topological persistence. This is defined as the density contrast of a critical pair (two critical points) and is chosen to be higher than a certain signal-to-noise (S/N) threshold. The S/N is typically defined relative to the root mean square persistence of a Poisson random density field. Spurious features or less significant ones are rejected such that only the topologically robust features contribute to the extracted cosmic web lattice. \\

\begin{figure}
        \raggedright{\includegraphics[width=.9\columnwidth]
        {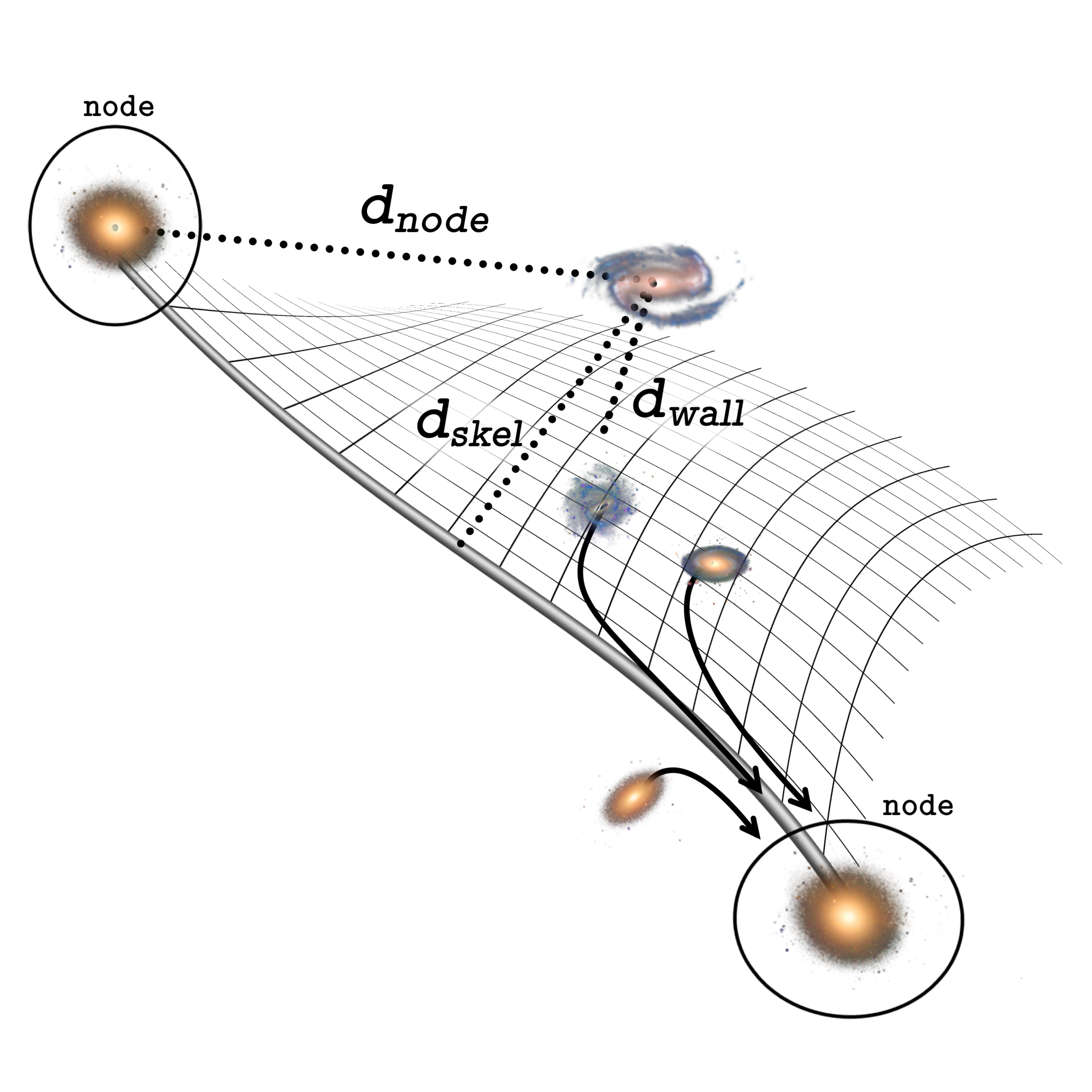}}
        \caption{\label{fig:Cosmic_Web_Sketch}Schematic view of the cosmic web metric. $d_{node}$, $d_{skel}$ and $d_{wall}$ are shown as dotted lines and represent the euclidean distance of a galaxy to the nearest node, filament and wall respectively.}
      \end{figure}

From the identified cosmic web features we define the cosmic web metric as a measure for large scale environment. An illustration of the distances as they were computed by \cite{Kraljic2018} is shown in Fig.~\ref{fig:Cosmic_Web_Sketch}. For each galaxy the euclidean distance to the nearest filament (wall) is measured as $d_{skel}$ ($d_{wall}$). Similarly, $d_{node}$ measures the 3D distance to the nearest node. Since {\small DisPerSE} does not account for the sampling of the data set, we normalize each distance by the redshift dependent mean inter-galaxy separation $\langle D_z \rangle$, which we compute from the number density $n(z)$ of galaxies in equidistant redshift bins: $\langle D_z \rangle = n(z)^{-1/3}$. For galaxies in SDSS DR 7, $\langle D_z \rangle$ ranges from $\sim 5 \, h^{-1}\textrm{Mpc}$ for nearby objects to $\sim 20 \, h^{-1}\textrm{Mpc}$ at the edge of the survey.

\subsection{The Working Sample}
\label{stn:The Working Sample}

\begin{figure*}
\centering
\includegraphics[width=1.04\columnwidth]{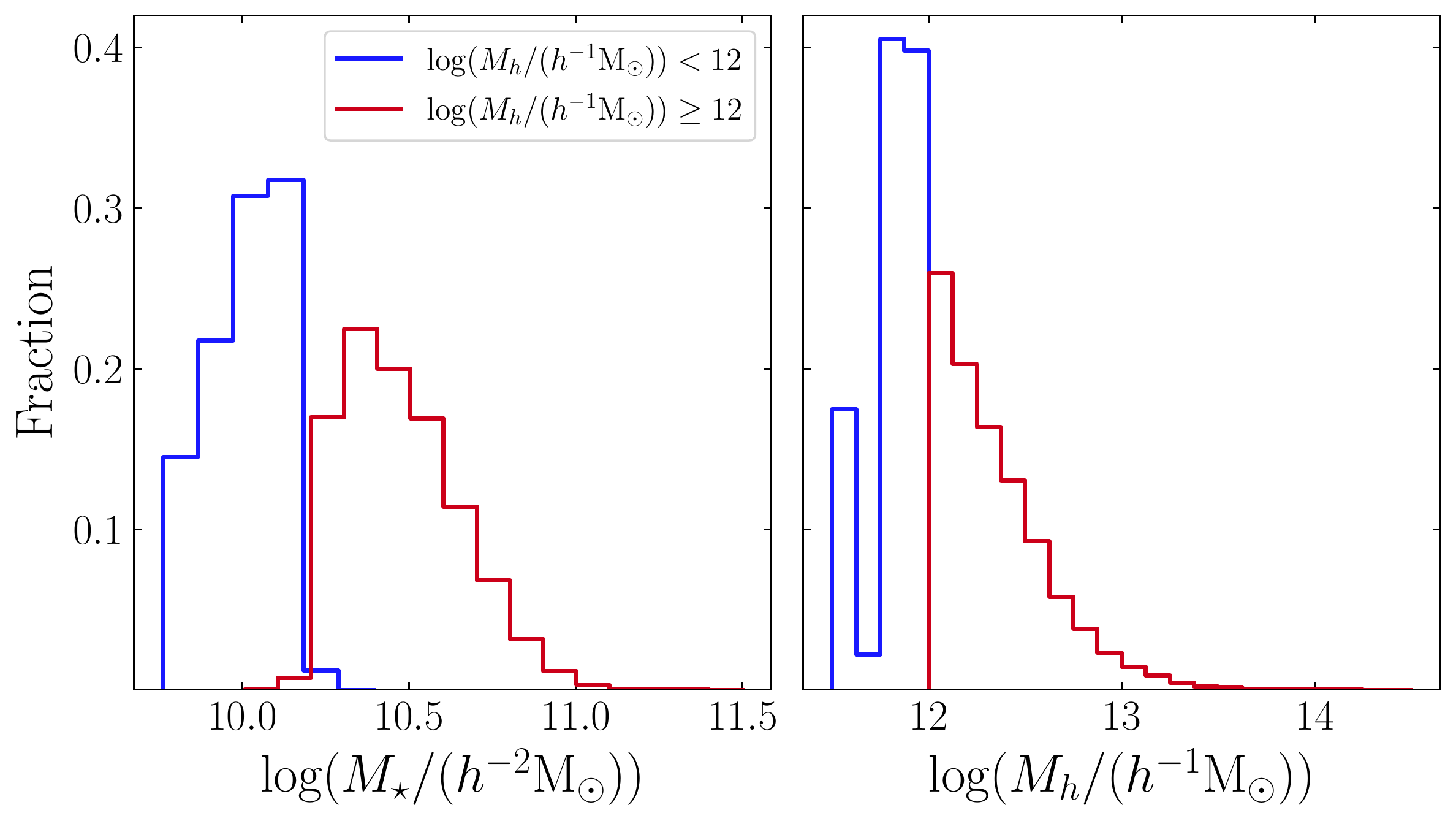}
\includegraphics[width=2\columnwidth]{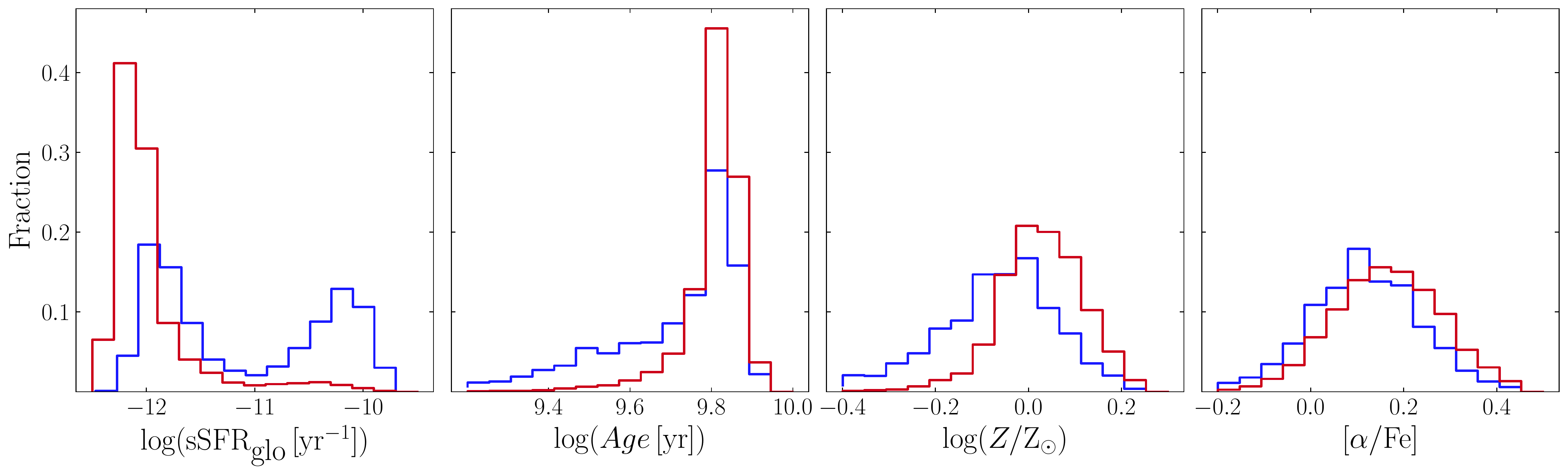}
\caption{\label{fig:Histograms}Top row: Weighted distribution of the sample centrals in stellar mass $M_\star$ and halo mass $M_h$.\\
Bottom row: From left to right, the panels show the weighted distribution of sample centrals in sSFR$_{\textrm{glo}}$, stellar age, metallicity and [$\alpha$/Fe].}
\end{figure*}

We extract central galaxies from \cite{Wang2014} whose $\textrm{sSFR}_\textrm{glo}$ is estimated by \cite{Brinchmann2004}. \cite{Gallazzi2021} compared the stellar masses from \cite{Bell2003} with other estimates and found an average scatter of $\sigma = 0.1\,$dex. Thus, we consider \emph{genuine} centrals those galaxies whose $M_\star$ is 3$\sigma$ higher than their associated satellites' mass.
We cross match the remaining galaxies with the catalogue of stellar ages, metallicities and [$\alpha$/Fe] values computed by \citet{Gallazzi2021} where we require a spectral S/N larger than 20. This ensures an uncertainty $<\, 0.2\,\textrm{dex}$ for the stellar ages and $<\, 0.3\,\textrm{dex}$ for the stellar metallicities, respectively.
Due to the small cosmological volume probed at at low redshifts and the lower number of galaxies at large redshifts respectively, we confine our analysis to $0.03 \leq z < 0.18$. This leaves us with a working sample of 69,193 centrals.\par\noindent
To correct for the Malmquist bias, we weight each galaxy by $1/V_{max}$, where $V_{max}$ corresponds to the comoving volume spanned by the maximum distance at which the galaxy would still be covered by the selection criteria of our sample. Furthermore, we weigh each galaxy by its spectral $w_{\textrm{S/N}}$ as determined by \citet{Gallazzi2021} in order to correct for incompleteness towards fainter galaxies at fixed $M_\star$ introduced by the cut in S/N. $w_{SN}$ is defined as the number ratio between the galaxies in \cite{Wang2014} and the galaxies in a subsample with S/N>20 in bins of stellar mass and $g-r$ colour.
Fig.~\ref{fig:Histograms} shows the weighted distribution of the sample centrals in $M_\star$, $M_h$ and the analysed properties. 
As expected from the tight $M_\star$-$M_h$ relation for central galaxies, those in haloes with masses above $10^{12}\, h^{-1}\textrm{M}_\odot$ are typically more massive than their counterparts in less massive haloes.
Furthermore, those in massive haloes are essentially quenched, older, metal richer and exhibit [$\alpha$/Fe] slightly larger than centrals in less massive haloes.\par\noindent
We note that the results presented in this work do not change qualitatively when we weigh by stellar mass $M_\star$ instead of $1/V_{max}$ or apply no weights at all. \\

As the stellar component is a reflection of the galaxy star formation history, $M_\star$ gauges the multitude of intrinsic processes.
We divide the sample into three stellar mass bins $\textrm{log}(M_\star / (h^{-2}\textrm{M}_\odot)) = [9.5,10),[10,10.5),[10.5,11.5]$, within which we assume that the influence of galaxy stellar mass on galaxy properties is substantially the same. 
Recent simulations by \cite{Hellwing2020} show that most of the haloes more massive than $10^{12} \, h^{-1} \textrm{M}_\odot$ reside in filaments and nodes while less massive haloes are more equally distributed among filaments, walls and voids \citep[c.f.][]{Kaiser1987}. In order to mitigate the impact such halo mass segregation may have on our results and preserve statistics, 
we also split our sample between $\textrm{log}(M_h / (h^{-1} \textrm{M}_\odot)) < 12$ and $\textrm{log}(M_h / (h^{-1} \textrm{M}_\odot)) \geq 12$. In the low halo mass regime 76 percent of the centrals have no satellite and are therefore largely isolated objects, whereas 81 percent of the centrals in haloes more massive than $10^{12}\, h^{-1}\textrm{M}_\odot$ host at least one satellite. In the following, we therefore refer to the former as field centrals and to the latter as group/cluster centrals respectively. Among the group/cluster centrals (75 percent of the sample), those with $M_h < 10^{13}\,h^{-1}\textrm{M}_\odot$ dominate our number counts (61 percent).\par\noindent
Table~\ref{tab:Working_Sample} shows how our sample is distributed across the chosen $M_\star - M_h$ bins.

\begin{table}
	\centering
	\caption{Number of sample centrals contained in the $M_\star$-$M_h$ bins specified in Section~\ref{stn:The Working Sample}}
	\label{tab:Working_Sample}
	\begin{tabular}{cccr} 
		\hline
		  & 
		  $\textrm{log}(M_h / (h^{-1} \textrm{M}_\odot )) < 12$ & $ \geq 12$\\
		\hline
		$\textrm{log}(M_\star / (h^{-2} \textrm{M}_\odot ))$ \\
		$[9.5,10)$ & 2,298 & 0 \\
		$[10,10.5)$ & 10,276 & 13,219 \\
		$[10.5,11.5]$ & 0 & 43,400  \\
		\hline
	\end{tabular}
\end{table}

\section{Methodology}
\label{stn:Methodology}

\subsection{Constraints on distances to the cosmic web features}
\label{stn:Constraints on distances to the cosmic web features}
By construction, the nodes as the high density peaks of the cosmic web are located at the intersection of filaments. Since we aim to measure the effect of the cosmic web outside of the nodes rather than the effect of the galaxy clusters residing at the nodes, we exclude the population of centrals with $d_{node} < 1\, \textrm{Mpc}$ for our analysis of $d_{node}$. This distance corresponds to the typical size of galaxy clusters \citep[e.g.][]{Hansen2005}. \par\noindent
Furthermore, the cosmic web nodes contaminate the mass gradients towards the filaments since galaxies close to the nodes are significantly more massive than those far away \citep{Kraljic2018}. This effect can be minimized by excluding galaxies that have smaller distances to the nodes than $d_{node}^{min}$. This minimum distance is chosen such that the gradients towards the nodes are disentangled from those towards the filaments while maintaining a large sample size for a statistically robust analysis of the galaxy properties. In order to estimate the contribution of the nodes to the mass gradients towards the filaments, we perform the same procedure described in \cite{Kraljic2018} where the distances to the filaments $d_{skel}$ are randomised in bins of $d_{node}$. By construction, only the contribution of the nodes remains in the mass gradients towards the reshuffled $d_{skel}$. For our sample, the contamination from the nodes can be reduced substantially when galaxies with $d_{node} \leq 1\,\textrm{Mpc}$ are rejected. During the following analysis of $d_{skel}$ we therefore only consider galaxies with larger distances to the nodes than $1\,\textrm{Mpc}/\langle D_z \rangle$, where $\langle D_z \rangle$ is the redshift-dependent mean inter-galaxy separation.\\
As for the filaments, we need to ensure that the galaxy property gradients towards the walls are not biased by galaxy mass segregation in the proximity of filaments and nodes. After randomising $d_{wall}$ in equidistant bins of $d_{skel}$, the contribution of the filaments is reduced for $d_{skel}^{min} = 1\,\textrm{Mpc}/\langle D_z \rangle$. Hence, we require the galaxies to be father away from the nodes \emph{and} filaments than $1\,\textrm{Mpc}/\langle D_z \rangle$ during the analysis of the galaxy properties with $d_{wall}$, which reduces our sample by 11,610 objects (16.7 percent).
For a more detailed explanation on how these sample-specific distances were computed we refer the reader to \citet{Kraljic2018}.\\

\subsection{Analysis of the Galaxy Properties}
\label{stn:Analysis of the Galaxy Properties}
Since the density gradients of the cosmic web are shallower than those of groups and clusters, their imprint on the average galaxy properties is expected to be relatively mild.
Furthermore, many galaxy properties correlate with galaxy stellar mass \citep[e.g.][]{Brinchmann2004, Gallazzi2005, Gallazzi2006} and several works suggest that galaxy evolution is primarily determined by galaxy stellar mass \citep[e.g.][]{Peng2010, Contini2019}. It is therefore crucial to keep $M_\star$ fixed in order to detect the second order effects from environment.\\

For each $M_\star - M_h$ bin, galaxies are divided into equidistant bins of $\textrm{log}(d_k / \langle D_z \rangle)$\footnote{The choice of the bin width is arbitrary. A sample specific value of $0.05\,\textrm{dex}$ has proven to provide high resolution while keeping a high number of galaxies contained in each bin.} where $d_k$ represents the distance to the cosmic web feature $k$ (i.e. nodes, filaments and walls). For each $\textrm{log}(d_k / \langle D_z \rangle)$ bin (which are indicated by the index $j$) that contains more than 10 galaxies, we compute the average $\bar{y}_{j,lin}$ of the galaxy property $y$, weighted by $w_i$ (see Section~\ref{stn:The Working Sample}).

\begin{equation}
    \bar{y}_{j, lin} = \frac{\sum_{i=1}^N w_{i,j} \times y_{i,j, lin}}{\sum_{i=1}^{N} w_{i,j}}.
        \label{eq:weighted_avg}
\end{equation}
Here, $w_{i}$ represents the weight of the $i^{th}$ galaxy (see Section~\ref{stn:The Working Sample}) and $N$ corresponds to the number of galaxies within each bin. 
In order to maintain the interpretation of the weighted average, we perform this calculation on a linear scale. However, the results presented in this work do not change qualitatively, if we compute the average property on logarithmic scale.
The weighted error of the mean is then computed as 
\begin{equation}
\label{eq:weighted_err_avg}
\sigma_{\bar{y}_{j,lin}} =  \sqrt{\frac{1}{N-1} \times \frac{\sum_{i=1}^N w_{i,j} \times (y_{i,j, lin} - \bar{y}_{j, lin})^2}{\sum_{i=1}^N w_{i,j}}}.
\end{equation}

Finally, we transform the linear average back to logarithmic scale and estimate the lower (upper) end of the uncertainty range by transforming $\bar{y}_{j, lin}- \sigma_{\bar{y}_{j, lin}}$ ($\bar{y}_{j,lin} + \sigma_{{\bar{y}_{j,lin}}}$) to logarithmic scale.

\subsection{Analysis of the Galaxy Property Offsets}
\label{stn:Analysis of the Galaxy Property Offsets}

We need to take into account that galaxies with small cosmic web distances have different average properties solely due to their higher average stellar mass. In order to control for this effect, we first compute the average relation between each property and $M_\star$. Due to the redshift dependence of the scaling relations, we define four redshift bins which are equally populated by our sample galaxies: $z= [0.03,0.06), [0.06, 0.078),[0.078,0.101), [0.101,0.18)$. Further, we use the $M_\star$ and $M_h$ bins specified in Section~\ref{stn:The Working Sample}. Each property-$M_\star$ relation is then computed for our sample centrals in each $z- M_\star - M_h$ bin separately by fitting a simple linear model $y^{rel} = a+b\times \textrm{log}(M_\star)$ to the linear galaxy properties, which takes into account the first order of the dependence on galaxy stellar mass.\\
For galaxies contained in a $z - M_\star - M_h - \textrm{log}(d_k / \langle D_z \rangle)$ bin we compute the weighted average values $\bar{y}_{j, lin}$ and $\bar{y}_{j,lin}^{rel}$ from the observed properties and the expected values from the scaling relation and redshift respectively. We define the average property offset in the galaxy property $y$ as $\Delta \bar{y}_{j} = \bar{y}_{j} - \bar{y}_{j}^{rel}$. This quantity gives us a measure of the average scatter of the scaling relation for property $y$ in a $z - M_\star - M_h - \textrm{log}(d_k / \langle D_z \rangle)$ bin.\\
By using the property offsets, we {\it i)} bypass the effect of stellar mass segregation between galaxies that are close to cosmic web features and those far away and {\it ii)} remove any dependence of galaxy properties on redshift, and {\it iii)} consider the entire sample to maximise the number statistics. 

\subsection{Measuring property gradients}
\label{stn:Measuring property gradients}

\begin{figure}
        \raggedright{\includegraphics[width=.9\columnwidth]
        {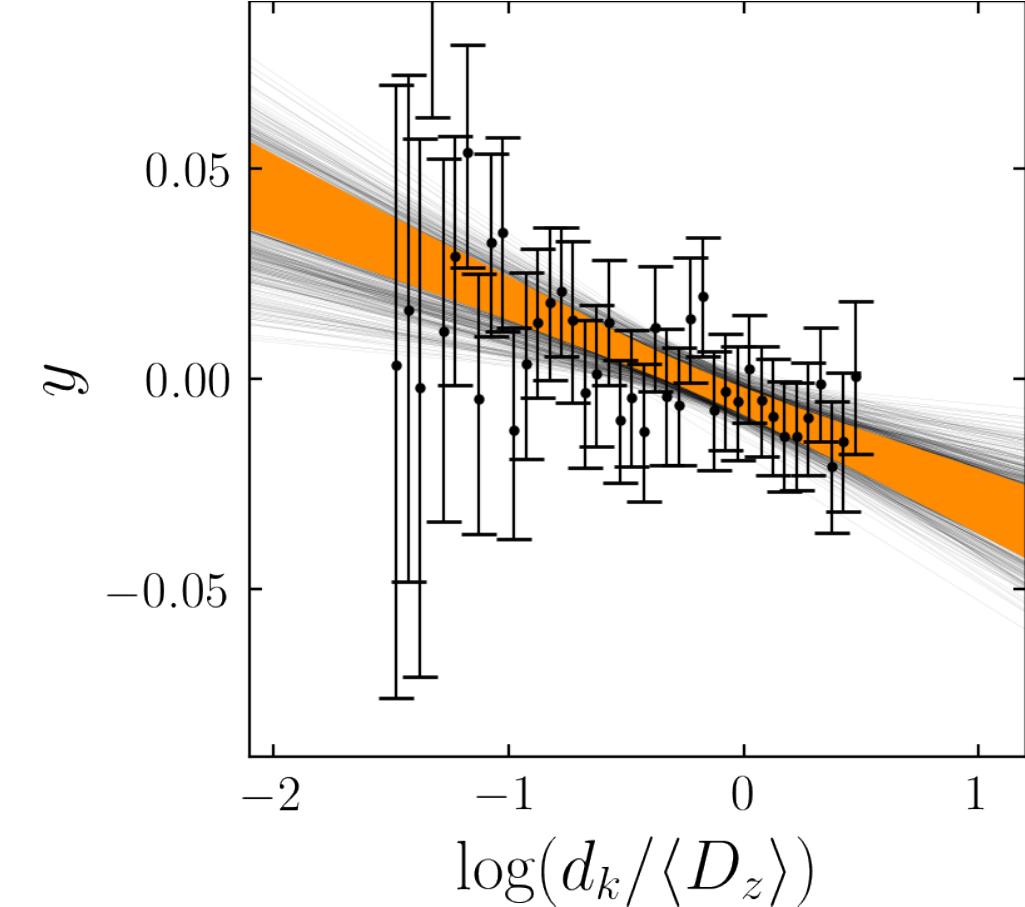}}
        \caption{\label{fig:Linmix}A visualisation of the fitting method which we use to quantify gradients with the cosmic web distances $\textrm{log}(d_k / \langle D_z\rangle)$. Here, y is an arbitrary observable whose averaged measurements $\bar{y}_j$ at different $\textrm{log}(d_k / \langle D_z\rangle)$ are represented by the orange data points. In our case, the $\bar{y}_j$ correspond to average galaxy property (or offsets) in a $M_\star$-$M_h$-$\textrm{log}(d_k / \langle D_z \rangle ) $-bin. Each thin black line represents a random draw from the posterior probability distribution. The thick orange line represents the best fit $\bar{y}_{fit}$ with the orange stripe indicating the $1\sigma$ confidence region.}
      \end{figure}
      
After performing the analysis described in the two previous sections, we derive the average galaxy properties $\bar{y}_j$ (and their offsets $\Delta \bar{y}_j$ ) in each $M_\star - M_h$ bin as a function of $\textrm{log}(d_k / \langle D_z \rangle)$.
In order to quantify the gradient $\nabla _k ^{\bar{y}}$ of the average property $\bar{y}$ w.r.t. $d_k$, we adopt the Bayesian method presented in \cite{Kelly2007} by performing a linear regression with a simple model \mbox{$\bar{y}^{fit} = a + \nabla _k ^{\bar{y}} \times \textrm{log}(d_k / \langle D_z \rangle)$} and similarly for the average property offsets \mbox{$\Delta \bar{y}^{fit} = a + \nabla _k ^{\Delta \bar{y}} \times \textrm{log}(d_k / \langle D_z \rangle)$} The precision only depends on the sampling of the galaxy distribution in redshift for which we account by weighting $d_k$ by the mean inter-galaxy separation $ \langle D_z \rangle$. The scatter in $d_k$ affects each normalised distance equally and is averaged out by the binning in $\textrm{log}(d_k / \langle D_z \rangle)$. We therefore assume that the uncertainty in $d_k$ does not affect our results both qualitatively and quantitatively, while we account for the uncertainties of the galaxy properties $\sigma_{\bar{y}_j}$.
Furthermore, we assume that the probability density functions of the independent variables $a$ and $\nabla$ can be approximated by a mixture of Gaussian functions. The method from \cite{Kelly2007} performs a Markov Chain Monte Carlo (MCMC) method of random draws of lines from the posterior probability distribution which is defined by the values $\bar{y}_j$ and their errors $\sigma_{\bar{y}_j}$. We select a minimum number of $n_{min}=500$ iterations for the MCMC to converge to the posterior. \par\noindent
From the drawn lines we extract the $16^{th}$, $50^{th}$ and $84^{th}$ percentile of their values at each point of $\textrm{log}(d_k / \langle D_z \rangle)$. The set of the $50^{th}$ percentiles represents the best linear fit, while the set of $16^{th}$ and $84^{th}$ percentiles mark the border of the $1\sigma$ confidence region respectively. 
We use sigma-clipping to identify outliers in the posterior distribution: if in the $j^{th}$ bin of $\textrm{log}(d_k / \langle D_z \rangle)$ the difference between $\bar{y}_j$ and the best fit is larger than $3 \sigma_{\bar{y}_{fit}}$, this value is excluded from the sample and the linear regression is performed again. Fig.~\ref{fig:Linmix} demonstrates the procedure where one can see that at large (low) values of $\textrm{log}(d_k / \langle D_z \rangle)$ the fit is obviously less well constrained than in regions that are thoroughly sampled by the $\bar{y}_j$. It is therefore important to probe a large range of $\textrm{log}(d_k / \langle D_z \rangle)$ in order to mitigate the uncertainty of the slope $\sigma _{\nabla}$ which is computed as the standard deviation of the $n$ slopes drawn from the posterior distribution. \\

In the analysis developed in Section~\ref{stn:Results}, we show the best fit only if {\it i)} there are more than five values $\bar{y}_j$ available and {\it ii)}  the gradient $\nabla$ (corresponding to the slope of $\bar{y}_{fit}$) is significantly different from zero at a $1 \sigma$ level. A complete list of the gradients measured in each property and in each $M_\star - M_h$ bin can be found in the Appendix~\ref{stn:appendix}.\\

\section{Results}
\label{stn:Results}

\subsection{Star Formation Rates}

\begin{figure}
        \raggedright{\includegraphics[width=.9\columnwidth]
        {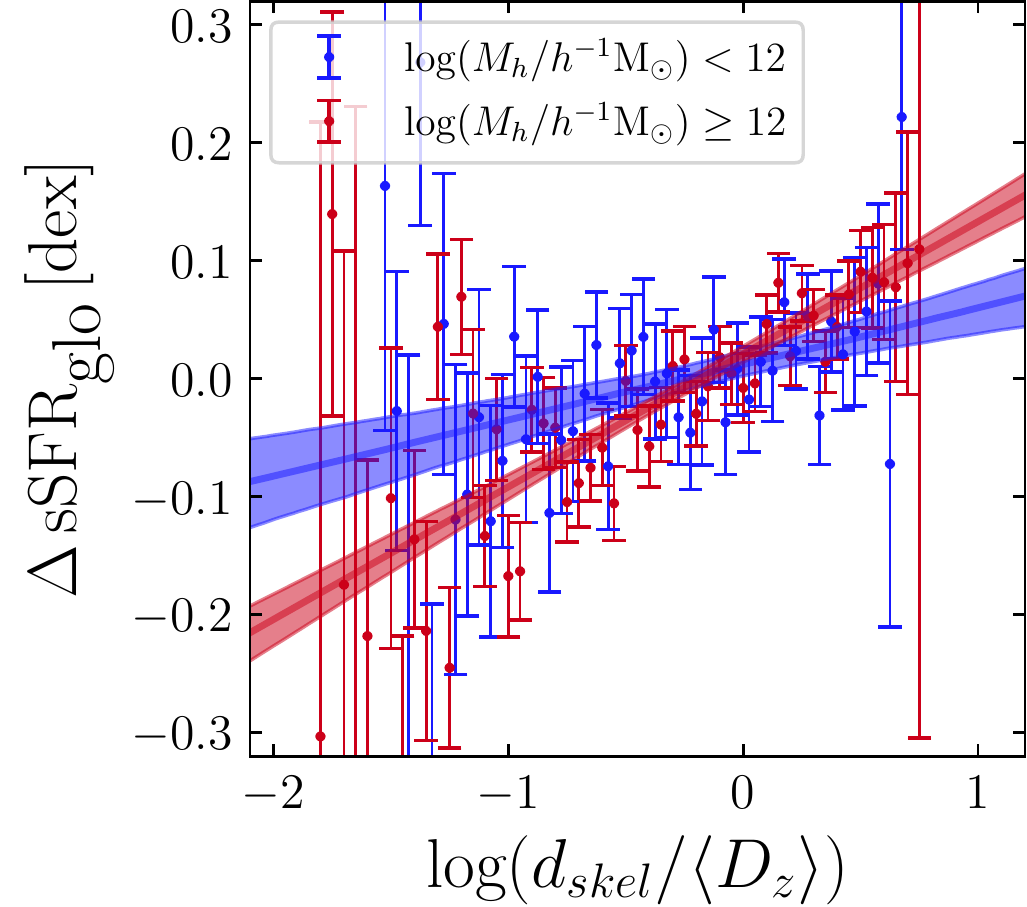}}
        \caption{\label{fig:sSFR} 
        Dependence of the average sSFR$_{\textrm{glo}}$ offsets ($\Delta {\textrm{sSFR}_{\textrm{glo}}}$) on the distances to the cosmic web filaments for SDSS central galaxies. Distances are normalised by the redshift-dependent mean inter-galaxy separation $\langle D_z \rangle$. Errorbars indicate the error of the mean for centrals contained in equidistant bins of $\textrm{log} (d_{skel}/\langle D_z \rangle)$ }. The gradients $\nabla _{skel}^ {\Delta \textrm{sSFR}_\textrm{glo}}$ can be found in Table~\ref{tab:Gradients_sSFR}. Both isolated centrals in lower mass haloes (blue) and centrals in groups (red) exhibit significantly increasing average sSFR increasing distance to the cosmic web filaments.
\end{figure}

It has been shown by \citet{Kraljic2018} that among galaxies in the GAMA survey there exists a segregation between the actively star-forming galaxies (sSFR$>10^{-10.8} \textrm{yr}^{-1}$) and quiescent galaxies, where star-forming galaxies are located at larger distance to the cosmic web filaments.
Further, they report that among the star forming population the median sSFR increases with increasing distance to the cosmic web filaments.\par\noindent
Our sample of SDSS central galaxies confirms a transition in sSFR from lower-than-average to higher-than-average with increasing distance to the cosmic web features.
The trends of the sSFR$_\textrm{glo}$-offset w.r.t. $d_{skel}$ in Fig.~\ref{fig:sSFR} show that in the proximity of the filaments at $\textrm{log}(d_{skel} / \langle D_z\rangle) = -1 $, centrals have lower-than-average  sSFR$_{\textrm{glo}}$ by $0.1\,\textrm{dex}$ ($\textrm{log}(M_h / (h^{-1}\textrm{M}_\odot)) < 12)$) and $\sim 0.13\,\textrm{dex}$ ($\textrm{log}(M_h / (h^{-1}\textrm{M}_\odot)) \geq 12)$. At the largest distances from the filaments ($\textrm{log}(d_{skel} / \langle D_z\rangle) = 1$), $\Delta \textrm{sSFR}_\textrm{glo}$ is higher than average by a similar amount. \par\noindent
Our sample centrals exhibit this behaviour also with increasing distance to the cosmic web nodes and walls (see Table~\ref{tab:Gradients_sSFR}).
The sSFR$_\textrm{glo}$ of centrals in low-mass haloes is systematically less affected by each of the cosmic web features than that of their counterparts in more massive haloes, although this difference is only significant for the cosmic web filaments ($2.4\sigma$) and walls ($2.5\sigma$). Field centrals ($M_h < 10^{12}\,h^{-1}\textrm{M}_\odot$) are more heavily impacted by $d_{node}$ and $d_{skel}$ than by $d_{wall}$.
For those in groups ($M_h \geq 10^{12} \,h^{-1}\textrm{M}_\odot$), the type of cosmic web feature does not systematically affect the magnitude of the $\Delta$sSFR$_{\textrm{glo}}$ gradients.
The typical sSFR$_\textrm{glo}$ variation over the sampled $d_{k}$ ranges is small ($\sim 0.2 \,\textrm{dex}$) compared to the full width at half maximum ($0.8\,\textrm{dex}$) of our sample's star-forming population ($\textrm{log(sSFR}_{\textrm{glo}} [\textrm{yr}^{-1}]) >-11 $, see Fig.~\ref{fig:Histograms}).

\subsection{Stellar Age}

\begin{figure*}
\centering
\begin{subfigure}{.6\columnwidth}
  \centering
  \includegraphics[width=\columnwidth]{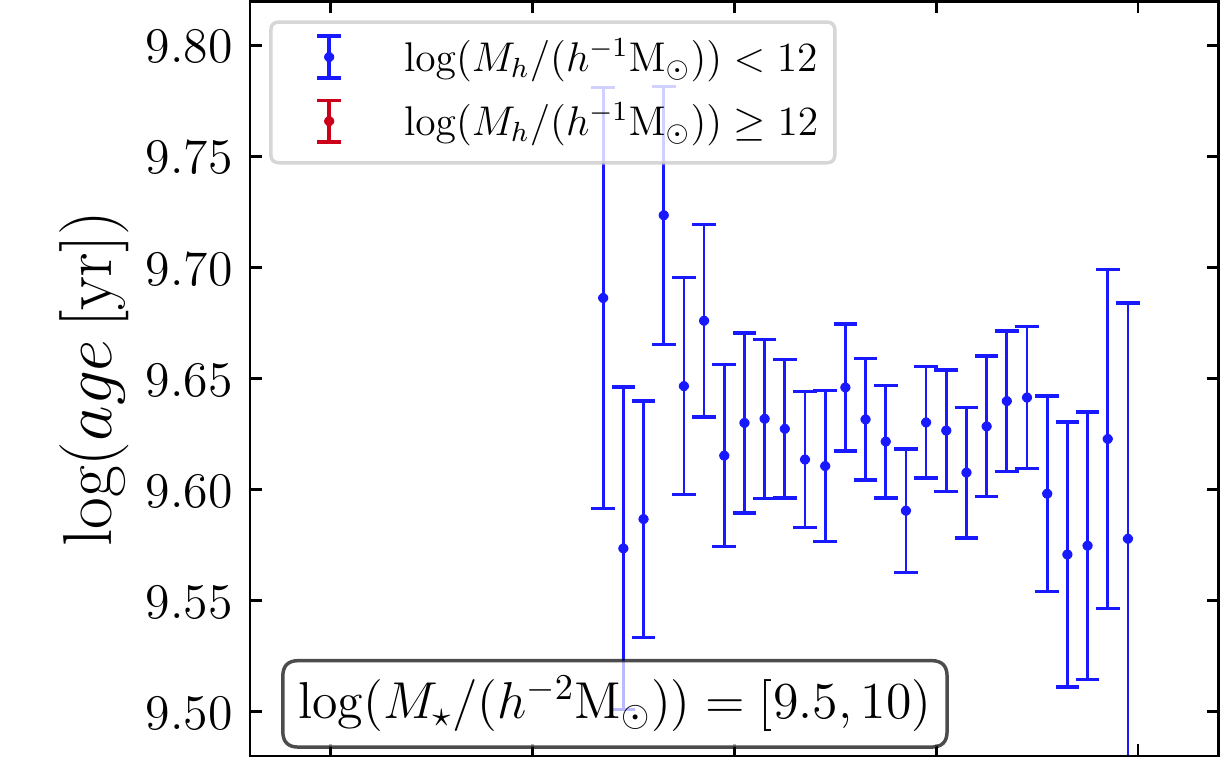}
\end{subfigure}%
\begin{subfigure}{.48\columnwidth}
  \centering
  \includegraphics[width=\columnwidth]{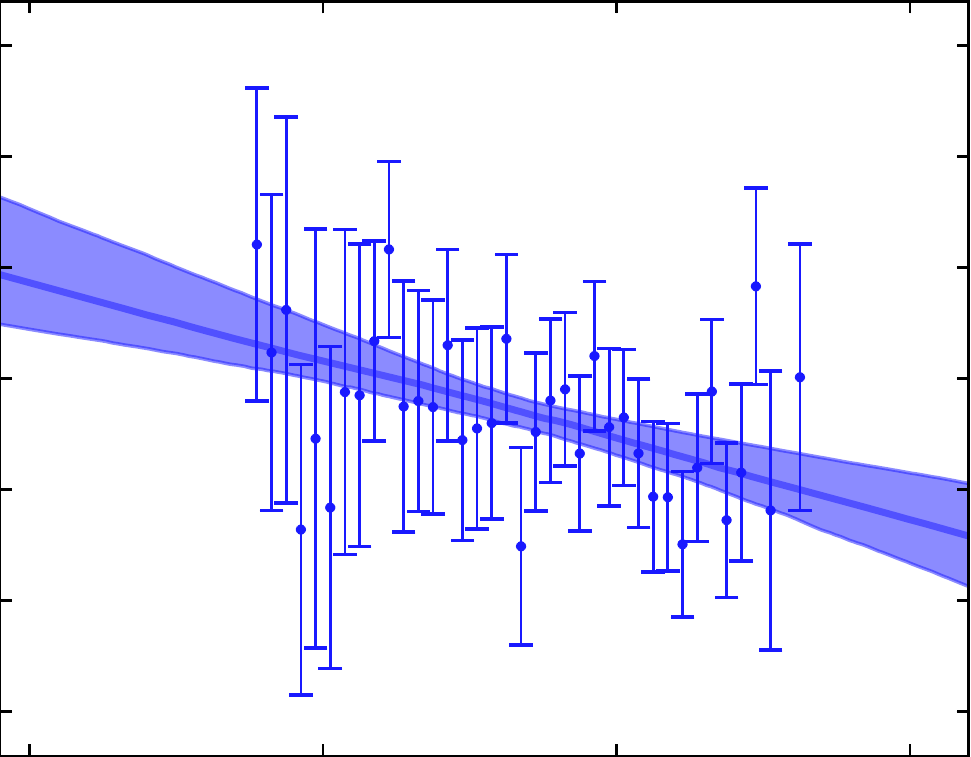}
\end{subfigure}%
\begin{subfigure}{.48\columnwidth}
  \centering
  \includegraphics[width=\columnwidth]{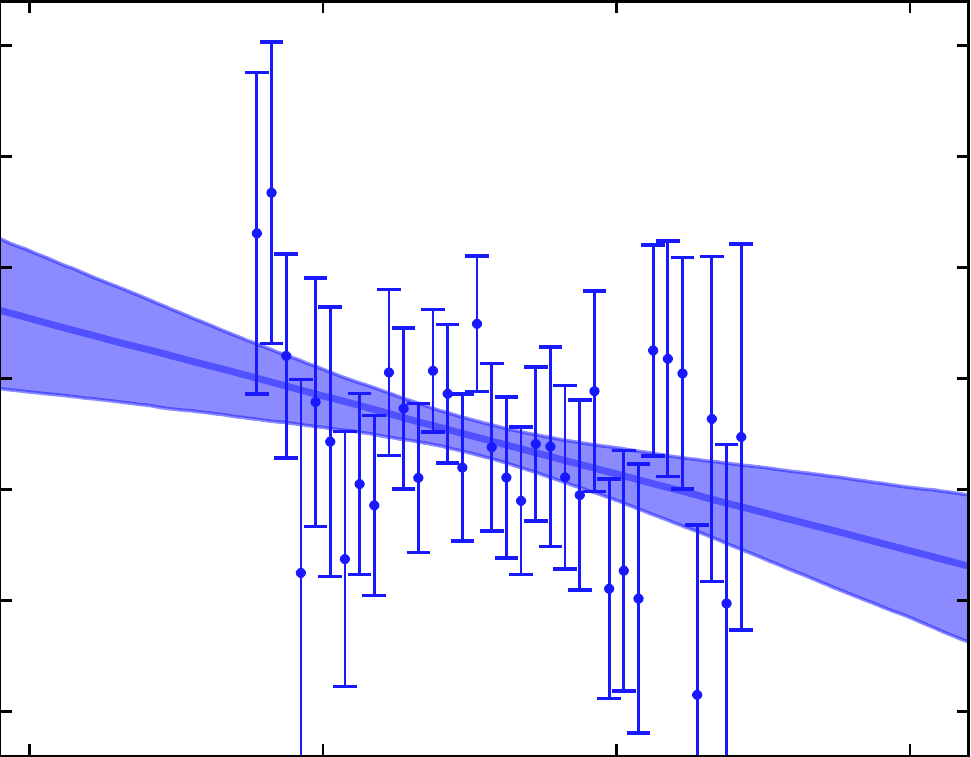}
\end{subfigure}

\vskip -2pt

\begin{subfigure}{.6\columnwidth}
  \centering
  \includegraphics[width=\columnwidth]{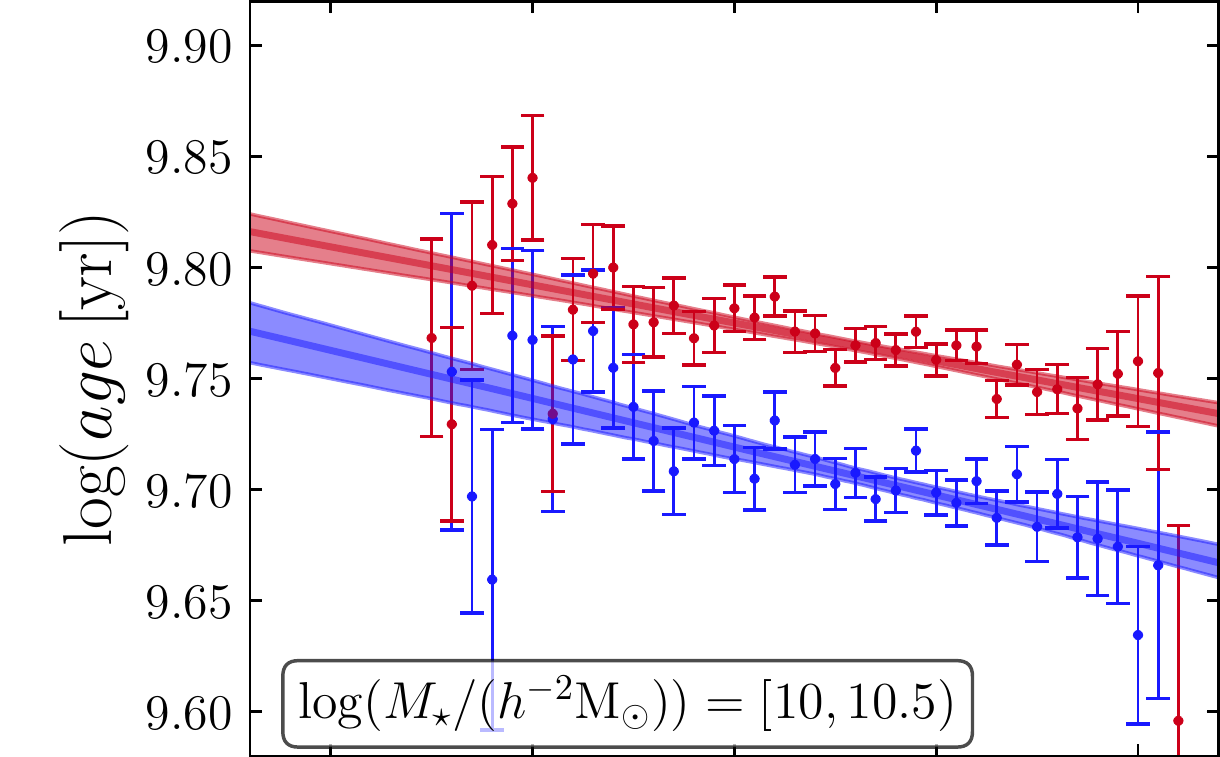}
\end{subfigure}%
\begin{subfigure}{.48\columnwidth}
  \centering
  \includegraphics[width=\columnwidth]{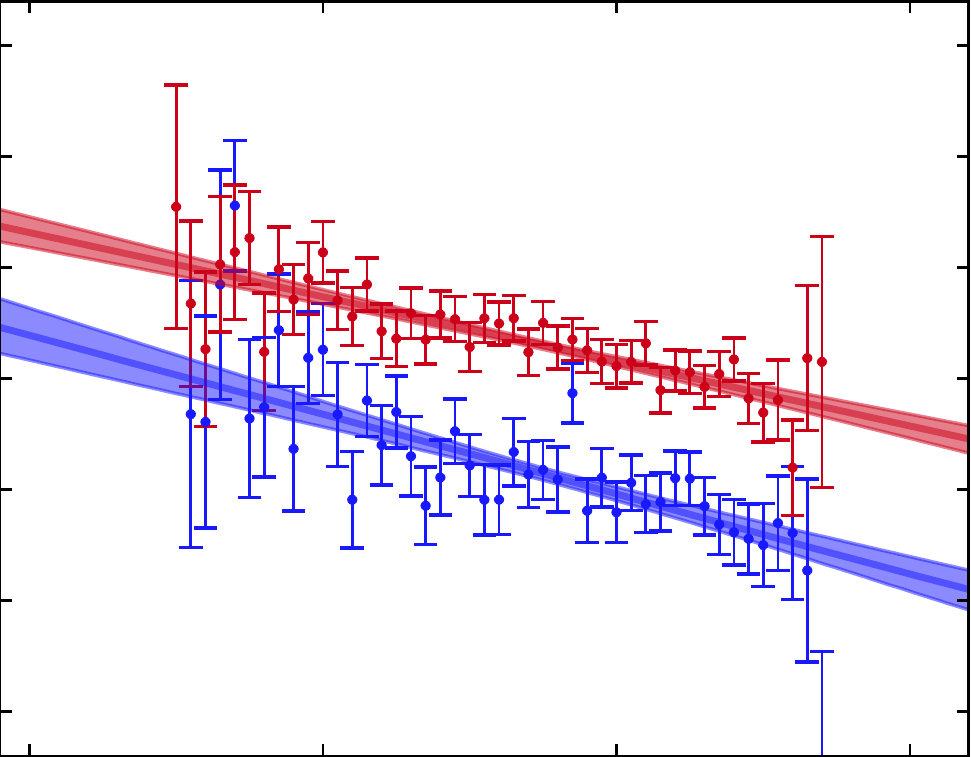}
\end{subfigure}%
\begin{subfigure}{.48\columnwidth}
  \centering
  \includegraphics[width=\columnwidth]{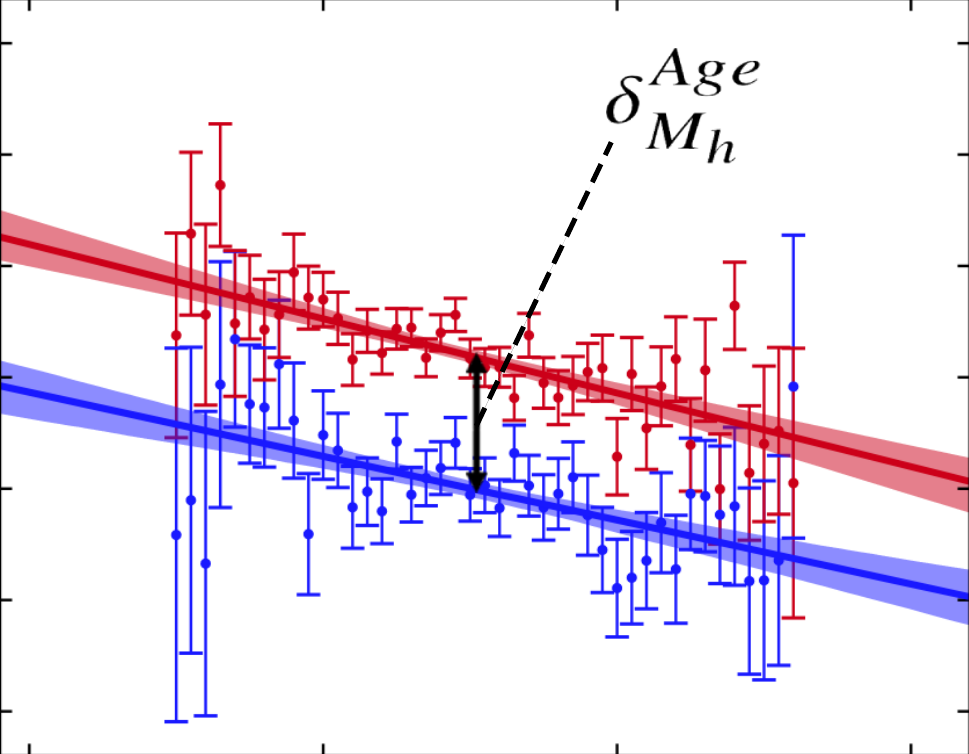}
\end{subfigure}

\vskip -2pt

\begin{subfigure}{.6\columnwidth}
  \centering
  \includegraphics[width=\columnwidth]{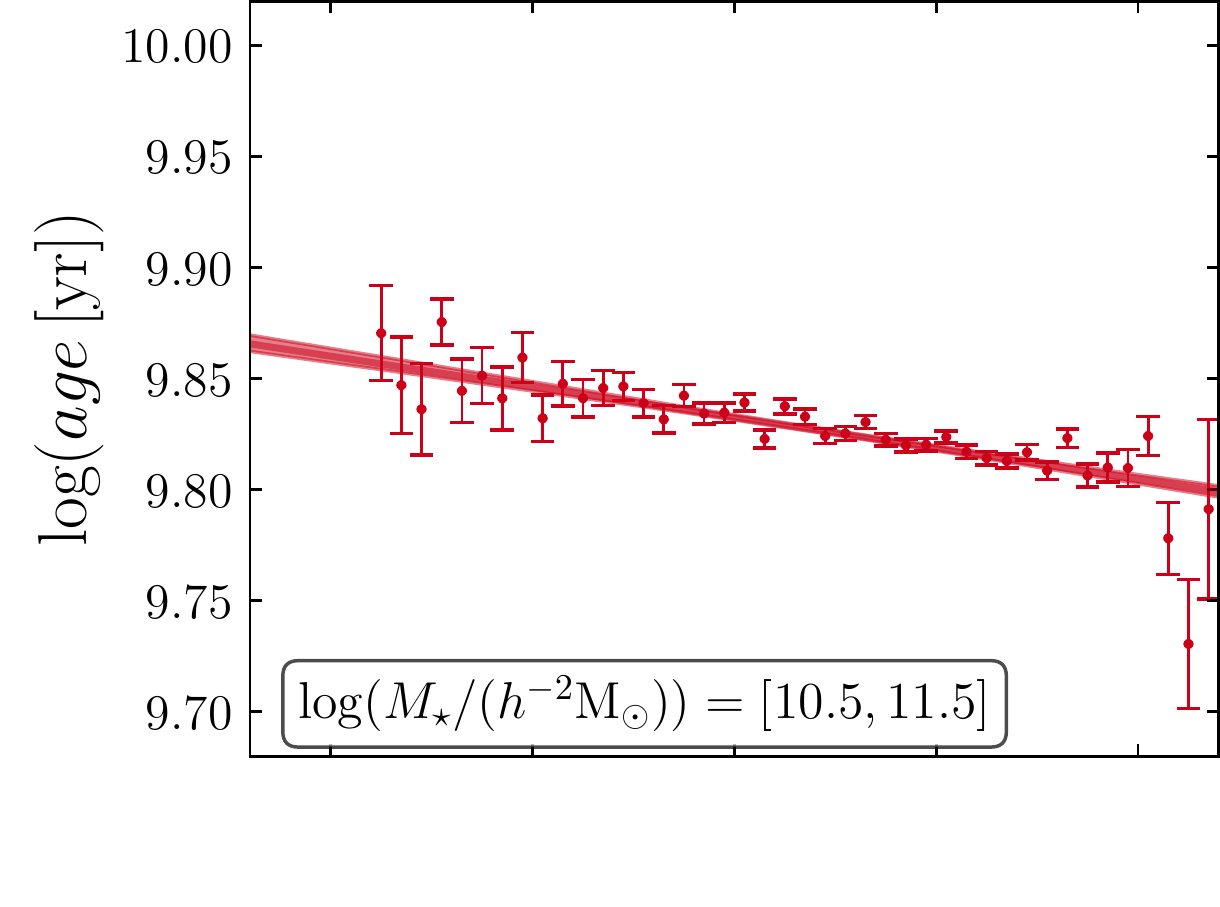}
\end{subfigure}%
\begin{subfigure}{.48\columnwidth}
  \centering
  \includegraphics[width=\columnwidth]{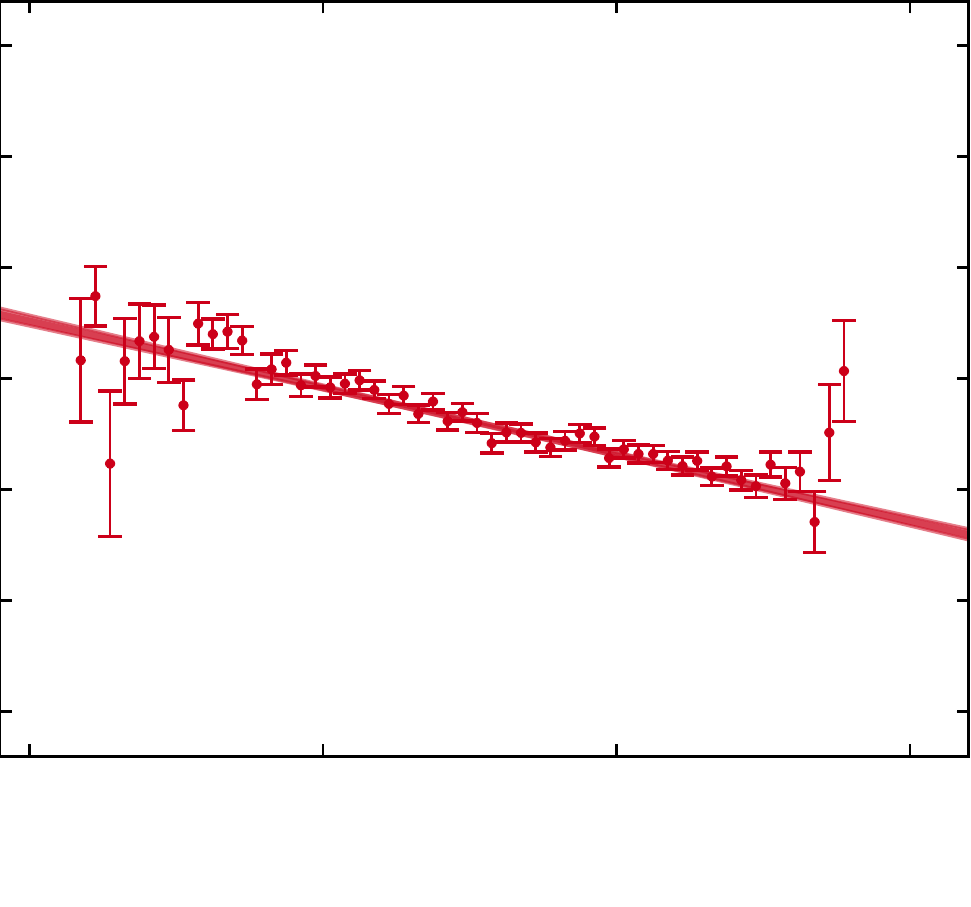}
\end{subfigure}%
\begin{subfigure}{.48\columnwidth}
  \centering
  \includegraphics[width=\columnwidth]{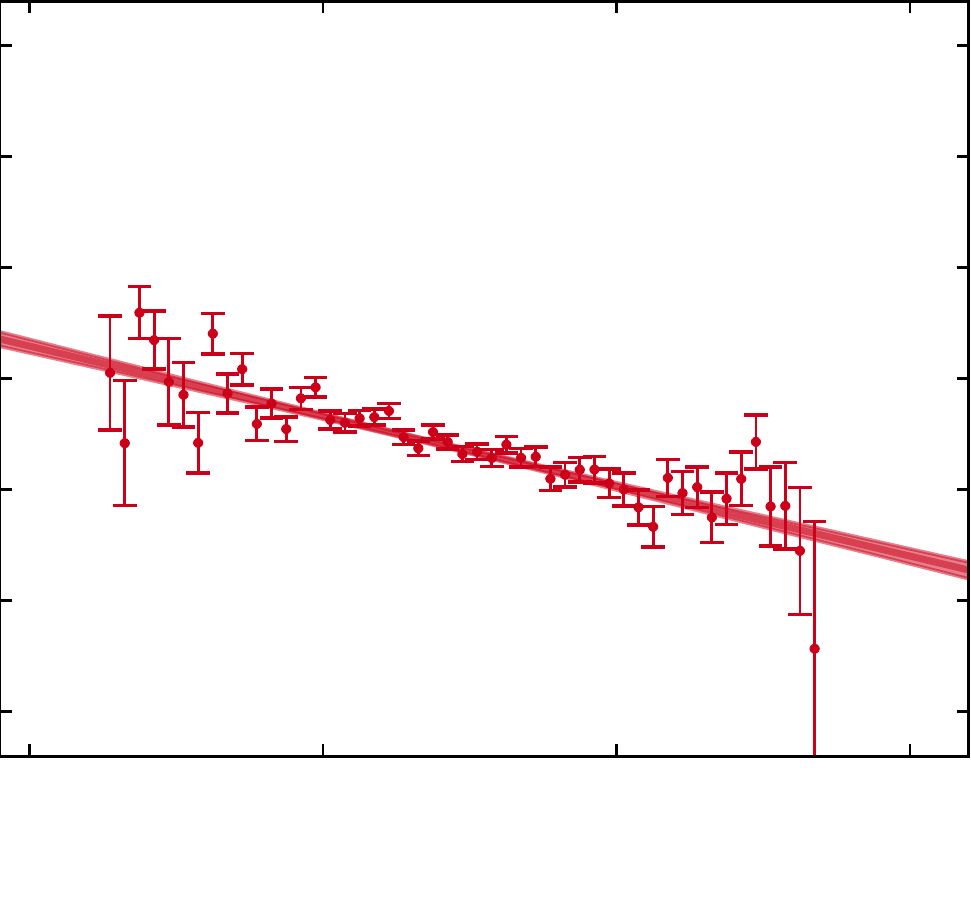}
\end{subfigure} 

\begin{subfigure}{.6\columnwidth}
  \centering
  \includegraphics[width=\columnwidth]{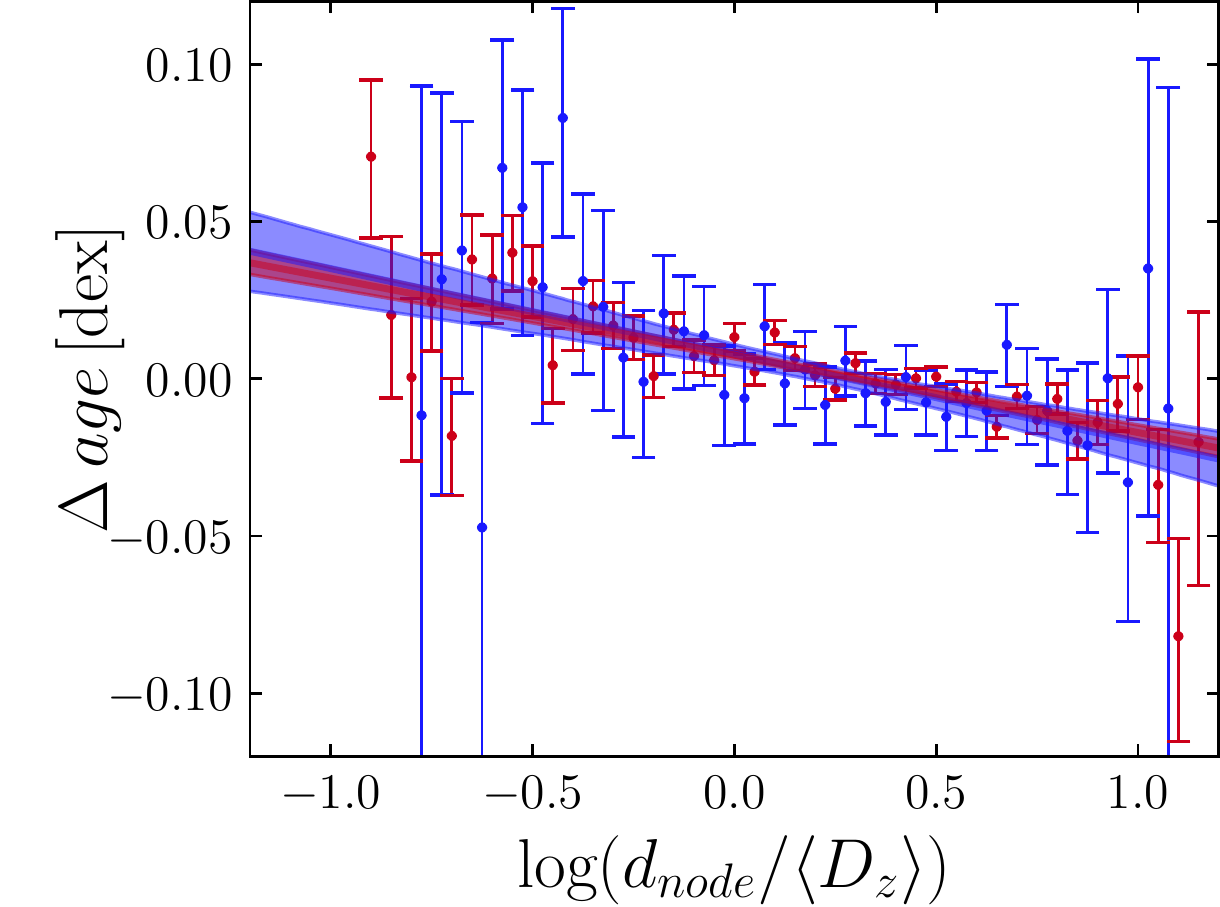}
\end{subfigure}%
\begin{subfigure}{.48\columnwidth}
  \centering
  \includegraphics[width=\columnwidth]{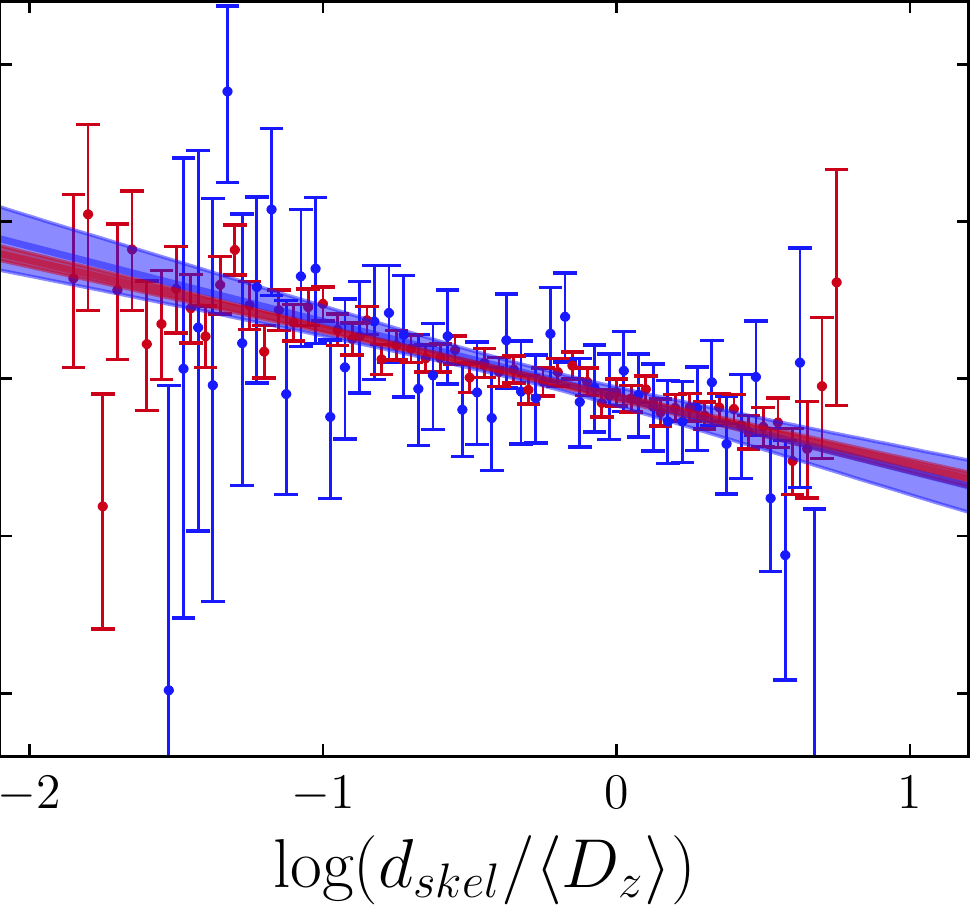}
\end{subfigure}%
\begin{subfigure}{.48\columnwidth}
  \centering
  \includegraphics[width=\columnwidth]{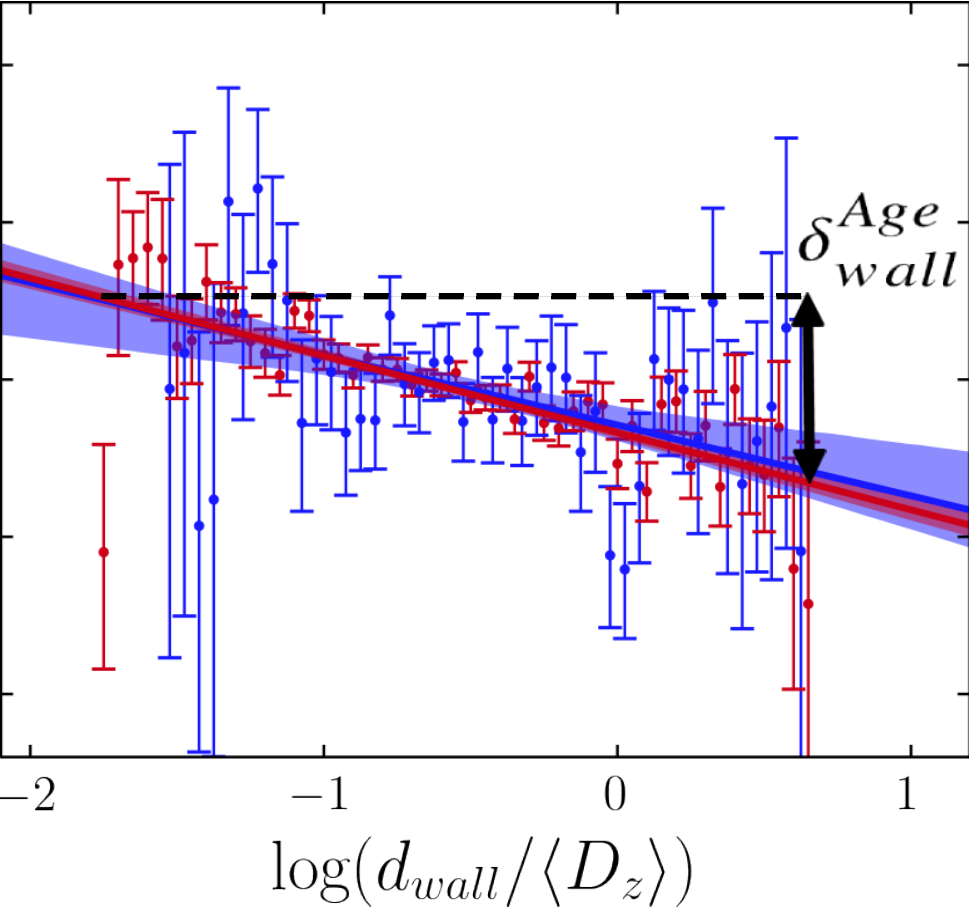}
\end{subfigure} \\

\caption{\label{fig:Age}Dependence of the weighted average stellar ages of sample central galaxies on the distances to the cosmic web features. In each panel the presentation of the results is similar to that shown Fig.~\ref{fig:Linmix}. The upper three panel rows correspond to three selected stellar mass bins, increasing from top to bottom. From left to right the trends for distance to the nodes, filaments and walls are shown respectively. The fourth row at the bottom shows the offset of the weighted average ages which were computed as described in Section~\ref{stn:Analysis of the Galaxy Property Offsets}. Colours indicate the two halo mass regimes. If the gradient that is measured as described in Section~\ref{stn:Measuring property gradients} is significant at the $1\sigma$ level (see Section~\ref{stn:Measuring property gradients}) we show the gradient as continuous line in the colour of the halo mass bin. The shaded area indicates the $1\sigma$ confidence region. The slope of the gradients are listed in Table~\ref{tab:Gradients_Age}. Within each stellar and halo mass bin central galaxies turn significantly younger with increasing distance to each of the cosmic web features.}
\end{figure*}

\begin{table*}
 	\centering
	\caption{From top to bottom: age gradient, transition distance, total age change over the sampled $d_k$ range (where $d_k$ is specific to the feature $k$, either node, filament or wall), median age change between field and groups.}
	\label{tab:Age}
	\begin{tabular}{lcccccr} 
        \hline
        \multicolumn{1}{c|}{} & & \multicolumn{1}{c}{$k = node$}
                    & \multicolumn{1}{c}{$skel$}  & \multicolumn{1}{c}{$wall$}\\
                                    
        \hline
        \multirow{1}{*}{$\nabla_{k} ^{\Delta age} \, [10^{-1} \textrm{dex/dex}]$}
     & field$^a$   
        & $-0.28 \pm 0.09$	
        & $-0.26 \pm 0.06$
        & $-0.21 \pm 0.09$
        \\
     & groups$^b$  
        & $-0.24 \pm 0.03$	
        & $-0.21 \pm 0.02$
        & $-0.25 \pm 0.02$
        \\
        \\
         \multirow{2}{*}{$\textrm{log}(d_{0,k}^{\Delta age} / \langle D_z \rangle) $ }
    &field    
        & $0.28 \pm 0.09$
        & $-0.24 \pm 0.06$
        & $-0.66 \pm 0.28$
        \\
    & groups 
        & $0.31 \pm 0.03$	
        & $-0.24 \pm 0.06$
        & $-0.67 \pm 0.05$
        \\
        \\
        \multirow{2}{*}{$\delta_{k}^{age}  \, [\textrm{dex}]$ }
     & field   & $0.075 \pm 0.017$
        & $0.051 \pm 0.018$
        & $0.050 \pm 0.009$
        \\
     & groups  & $0.064 \pm 0.008$	
        & $0.055 \pm 0.008$
        & $0.060 \pm 0.009$
        \\
        \\
        \multirow{1}{*}{$\delta_{M_h}^{age} \, [\textrm{dex}]$}
     &  & $0.059 \pm 0.005$
        & $0.054 \pm 0.005$
        & $0.061 \pm 0.005$
        \\        
		\hline
	\end{tabular}
\\
$^a$: in host haloes with $\textrm{log}(M_h / (h^{-1}\textrm{M}_\odot)) < 12$\\
$^b$: in host haloes with $\textrm{log}(M_h / (h^{-1}\textrm{M}_\odot)) \geq 12$
\end{table*}

In contrast to the global sSFR which \citet{Brinchmann2004} estimated for the whole galaxy optical body, the stellar population properties adopted from \citet{Gallazzi2021} were computed for the inner $3\,\textrm{arcsec}$ diameter fibre spectra with spectral synthesis modelling. Thus, the analysis of the stellar ages, metallicities and element abundance ratio [$\alpha$/Fe] provides an independent measure of how cosmic web environment affects galaxy star formation history.

The first three panels (from top to bottom) of Fig.~\ref{fig:Age} show the weighted average stellar ages as a function of distance to the cosmic web features in the three selected stellar mass bins. Each row has three panels that share the same ordinate. These panels correspond to the distance to the cosmic web nodes ($d_{node}$), filaments ($d_{skel}$) and walls ($d_{wall}$) respectively. Finally, the fourth row at the bottom of Fig.~\ref{fig:Age} shows the average stellar age offsets of our sample centrals that were computed as described in Section~\ref{stn:Analysis of the Galaxy Property Offsets}) The two halo mass regimes are colour-coded in red and blue for $M_h < 10^{12} \, h^{-1}\textrm{M}_\odot$ and $M_h \geq 10^{12} \, h^{-1}\textrm{M}_\odot$ respectively.\par\noindent
We start by first considering the trends of average stellar ages with distance to the cosmic web features. The tight $M_\star$-$M_h$-correlation that centrals follow leaves us with no centrals in massive haloes up to $M_\star = 10^{10}\,h^{-2}\textrm{M}_\odot$ and zero centrals in low-mass haloes above $M_\star = 10^{10.5}\,h^{-2}\textrm{M}_\odot$. In general, central galaxies become on average younger with increasing distances to the nodes, filaments and walls. While for the least massive centrals this behaviour is associated with a high uncertainty due to the small number of objects which cover a smaller range of $d_k$, the trends are better constrained at larger stellar masses. \par\noindent
In the intermediate stellar mass bin, the decrease of the average stellar ages with increasing distance is significant in each $M_h$ bin. Considering the uncertainties, the gradients do not depend on halo mass or the type of cosmic web feature (see Table~\ref{tab:Gradients_Age}). In this stellar mass bin both halo mass regimes are fairly equally represented (44 percent with $M_h < 10^{12}\, h^{-1}\textrm{M}_\odot$, see Table~\ref{tab:Working_Sample}) which allows us to estimate the effect of $M_h$ alone. Between centrals that reside in differently massive haloes, we observe an age-offset across the full $d_k$ range: centrals in massive haloes exhibit systematically higher average ages than those in low-mass haloes, in agreement with the findings of \cite{Pasquali2010} and \citet{Gallazzi2021}. Based on this fairly constant offset, let us define the typical age difference between the two halo mass regimes $\delta_{M_h}^{age}$ as the median age difference between the two halo mass regimes at fixed stellar mass. This quantity is visualised as a black arrow in the second row, third panel of Fig.~\ref{fig:Age} and will help us later to compare the impact of the central's halo mass with that of the individual cosmic web features. By construction, $\delta_{M_h}^{age}$ does not depend on $d_k$. The small differences of its values listed in Table~\ref{tab:Age} stem from the slightly different samples used for the analysis of the $d_k$ (see Section~\ref{stn:Constraints on distances to the cosmic web features}).

Because of the non-negligible width of the $M_\star$ - $M_h$ bins, the age-gradients shown in the top panels of Fig.~\ref{fig:Age} could be biased by the mass segregation towards the cosmic web features within the narrow range of the $M_\star$ bins. In order to control for this effect, we compute the average age offset as described in Section~\ref{stn:Analysis of the Galaxy Property Offsets} and show them in fourth row at the bottom of Fig.~\ref{fig:Age}.
For both halo mass regimes, we observe that central galaxies close to nodes, filaments and walls are typically older than the average population of central galaxies, whereas centrals at large distances are significantly younger. This gradient is most pronounced for centrals in low-mass haloes towards the nodes where the gradient amounts to $\nabla _{node} ^{\Delta age} = -0.033 \pm 0.006 \, \textrm{dex/dex}$.
Moving away from the nodes we also see centrals in more massive haloes turning from older-than-average to younger-than-average. Considering the uncertainties, the gradients of the age offset $\nabla _k^{\Delta age}$ towards the filaments ($k=skel$) and walls ($k=wall$), respectively, are consistent with the gradient towards the nodes $\nabla _{node}^{\Delta age}$ (see Table~\ref{tab:Age}). 
Between field and group central galaxies, we only observe a significantly different $\Delta age$ gradient towards the nodes, where the gradient of the field centrals is steeper (at a 1.9$\sigma$ level).

The second row of Table~\ref{tab:Age} lists the `transition distance' $\textrm{log}(d_{0,k}^{age} /\langle D_z \rangle)$, which we define as the distance to the cosmic web feature $k$ at which centrals turn from older-than-average to younger-than-average. 
Interestingly, the cosmic web features appear to have a similarly strong imprint on the average age offsets of central galaxies (as quantified by the gradient of the property $\nabla _k^{\Delta age}$) despite having different transition distances. For both field and group centrals, the transition distance increases by nearly $1\, \textrm{dex}$ as we move from the walls to the filaments and nodes. 

In order to compare the impact of the cosmic web environment with that of the central's halo mass, we define $\delta_{k}^{age}$ as the typical variation of the average age offset across the sampled $d_{k}$ range. 
In the bottom row, third panel of Fig.~\ref{fig:Age}, $\delta_{wall}^{age}$ is visualised by the black arrow.
$\delta_{k}^{age}$ is largest for group centrals that approach the cosmic web nodes. Here the variation over the sampled $d_{node}$ range is $\delta_{node}^{age} = 0.075 \pm 0.017\, \textrm{dex}$, corresponding to $1.3 \pm 0.3 \, \textrm{Gyr}$ at the median stellar age of the sample. 
Since the sampled $d_{k}$ range varies between the cosmic web features and the different halo mass regimes, we note that the $\delta_{k} $ must only be compared with $\delta_{M_h}$, while the gradients $\nabla _k$ should be used to compare the cosmic web features $k$ amongst each other.\par\noindent
Although the variation of $\Delta age$ induced by the cosmic web features $\delta_{k}^{age}$ suffer from larger scatter, their median is consistent with $\delta_{M_h}^{age}$ (see Table~\ref{tab:Age}). The typical value of \mbox{$\delta_{M_h}^{age} = 0.055 \pm 0.005 \, \textrm{dex}$}, corresponds to $0.9 \pm 0.1\, \textrm{Gyr}$ at the median stellar age of the sample centrals. 
This indicates that the distance to each cosmic web feature has a similarly heavy impact on the average ages as the difference between the two halo mass regimes. However, for centrals in low-mass haloes, $d_{node}$ has a slightly higher (at a 1$\sigma$ level) influence on the average ages than $M_h$.

\subsection{Metallicity}
\label{stn:Metallicity}

\begin{figure*}

\centering
\begin{subfigure}{.6\columnwidth}
  \centering
  \includegraphics[width=\columnwidth]{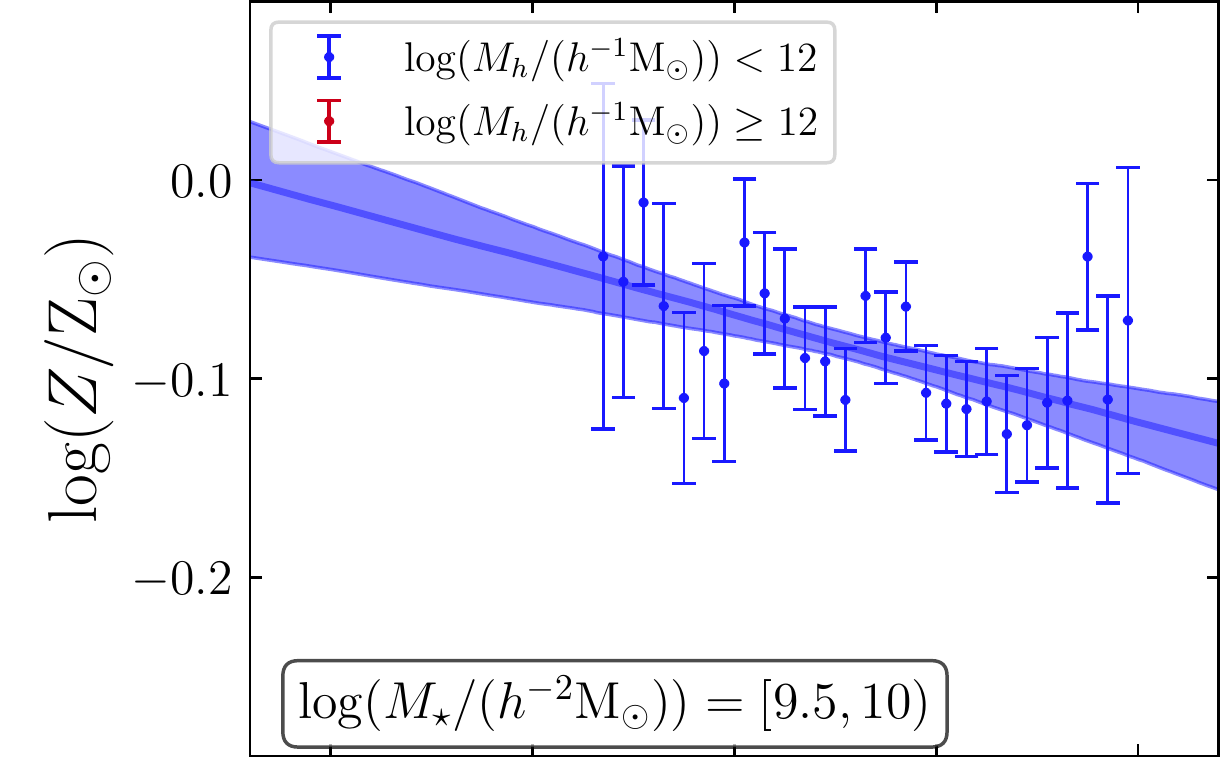}
\end{subfigure}%
\begin{subfigure}{.48\columnwidth}
  \centering
  \includegraphics[width=\columnwidth]{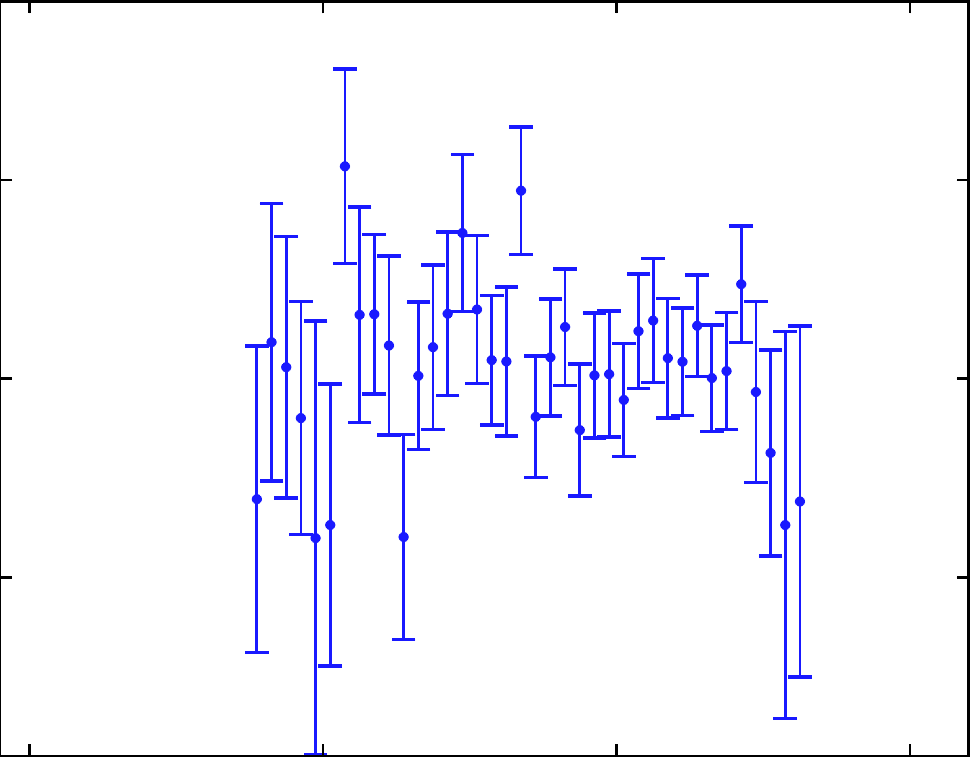}
\end{subfigure}%
\begin{subfigure}{.48\columnwidth}
  \centering
  \includegraphics[width=\columnwidth]{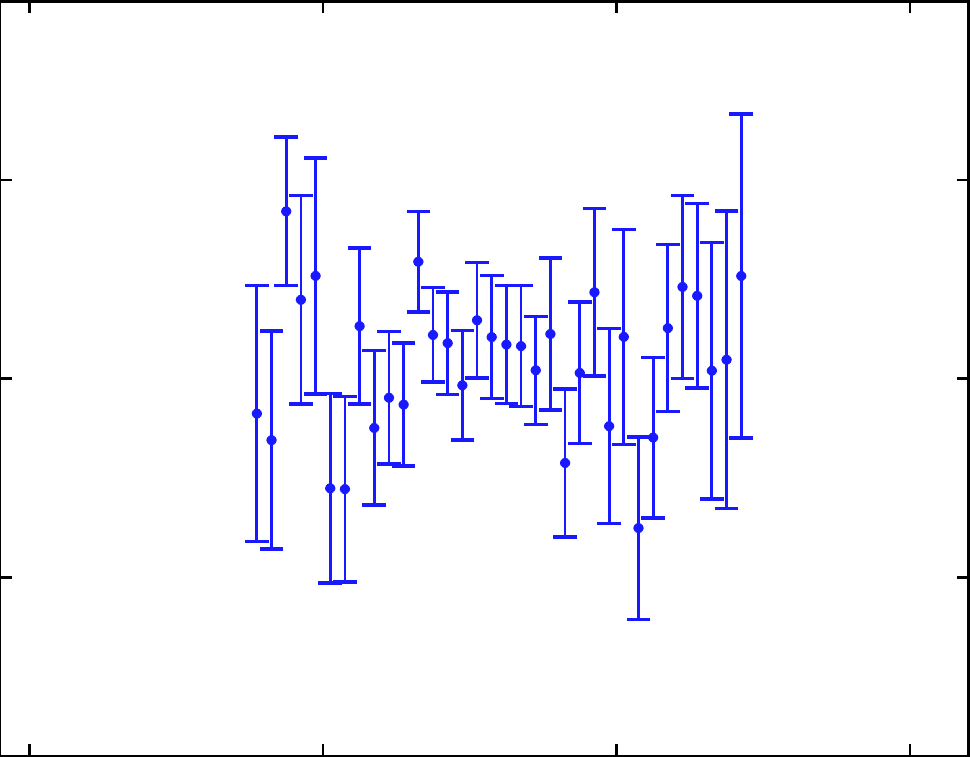}
\end{subfigure} 

\vskip -2pt

\begin{subfigure}{.6\columnwidth}
  \centering
  \includegraphics[width=\columnwidth]{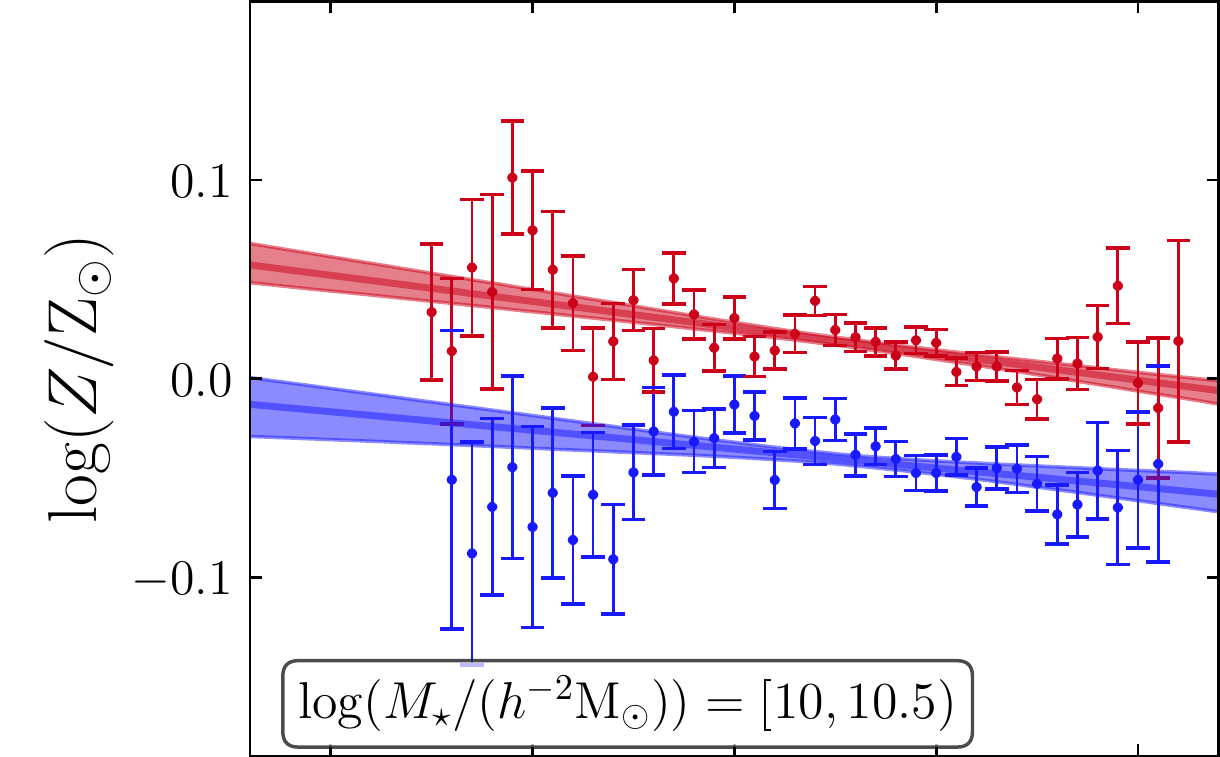}
\end{subfigure}%
\begin{subfigure}{.48\columnwidth}
  \centering
  \includegraphics[width=\columnwidth]{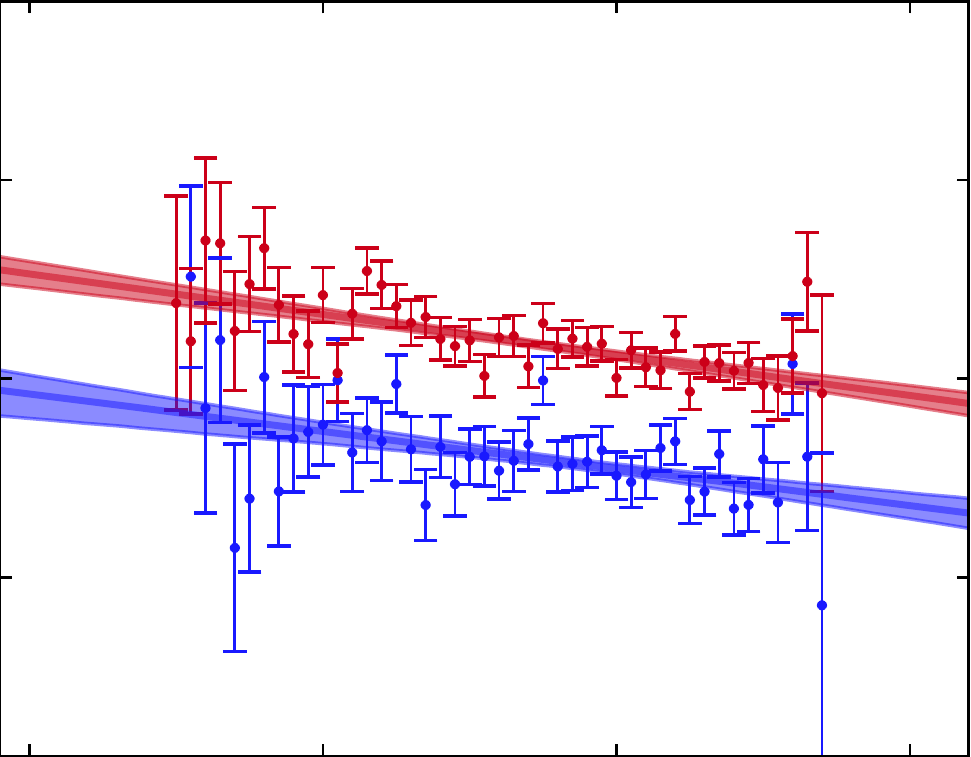}
\end{subfigure}%
\begin{subfigure}{.48\columnwidth}
  \centering
  \includegraphics[width=\columnwidth]{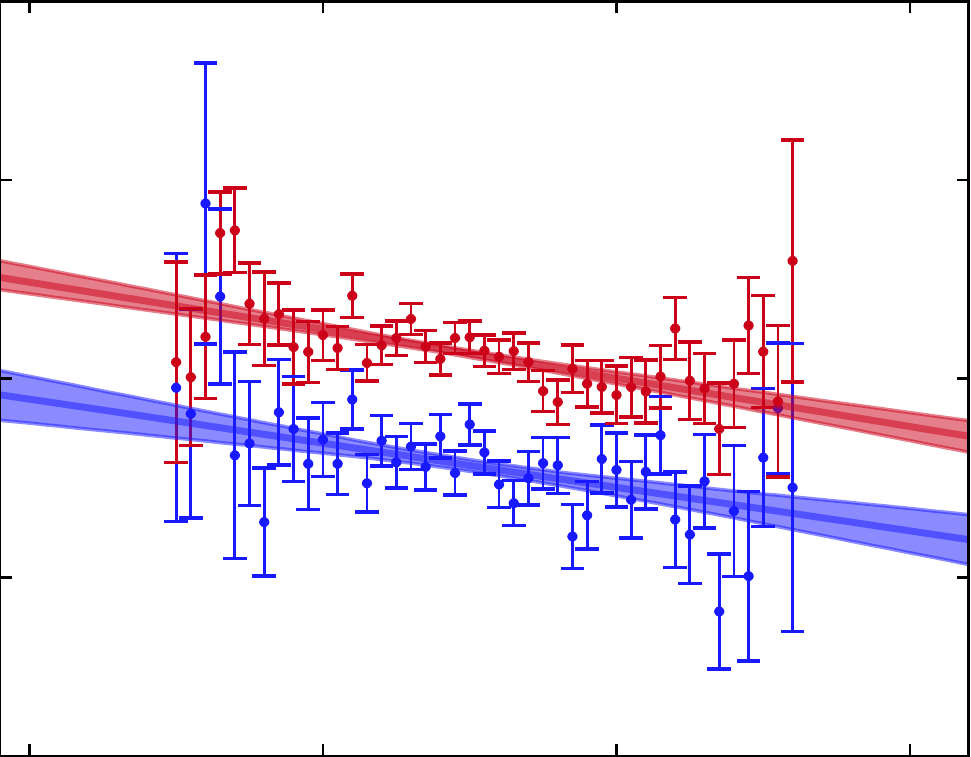}
\end{subfigure}

\vskip -1pt

\begin{subfigure}{.6\columnwidth}
  \centering
  \includegraphics[width=\columnwidth]{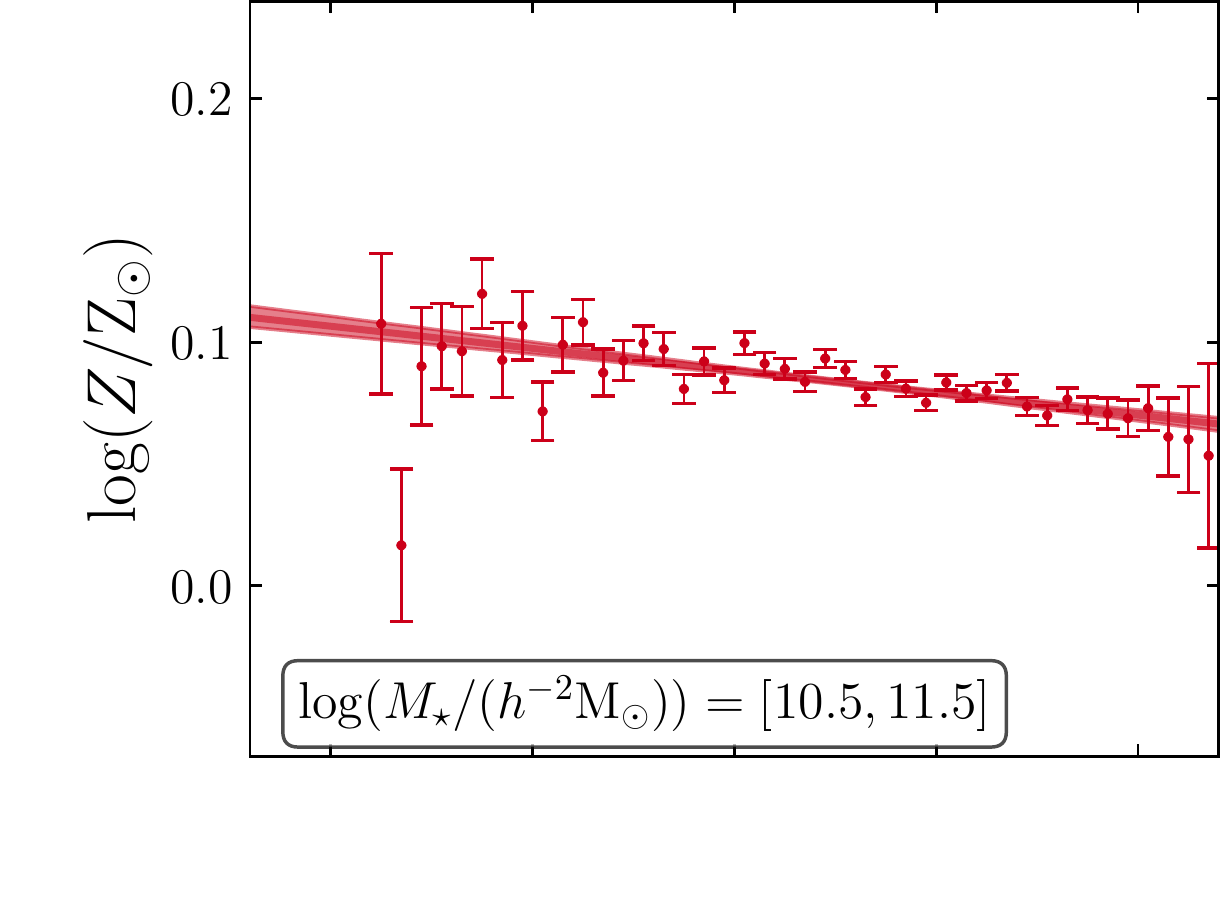}
\end{subfigure}%
\begin{subfigure}{.48\columnwidth}
  \centering
  \includegraphics[width=\columnwidth]{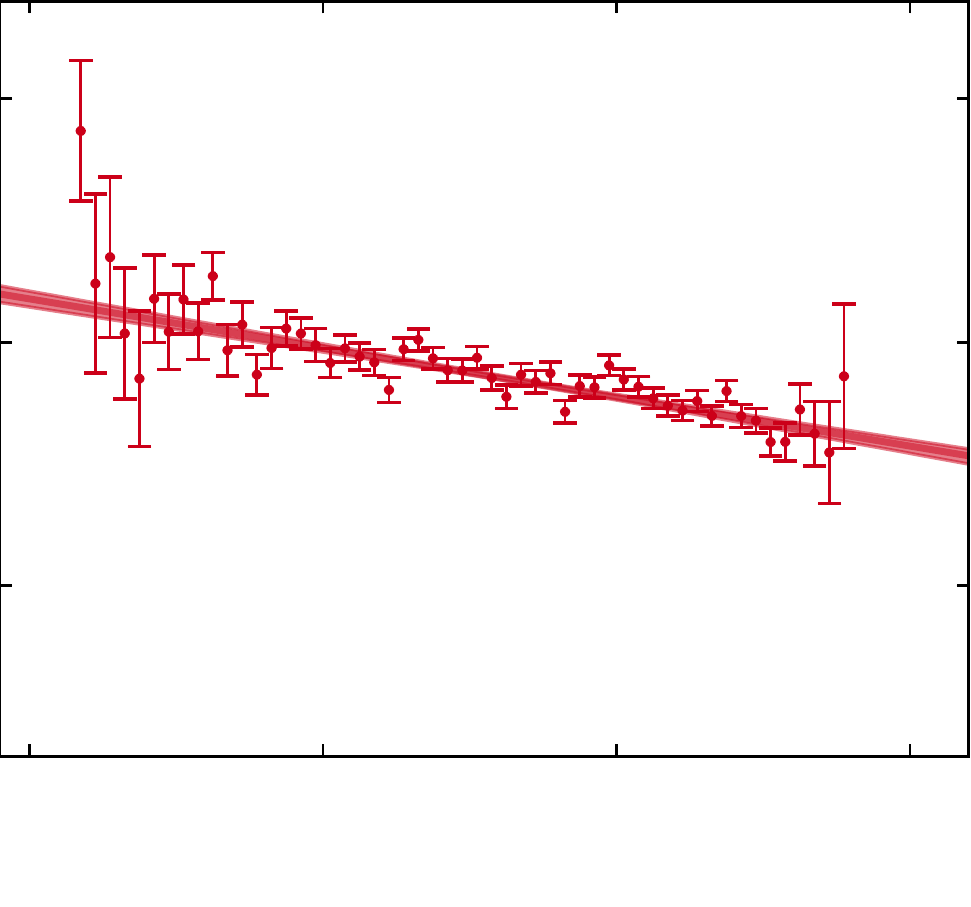}
\end{subfigure}%
\begin{subfigure}{.48\columnwidth}
  \centering
  \includegraphics[width=\columnwidth]{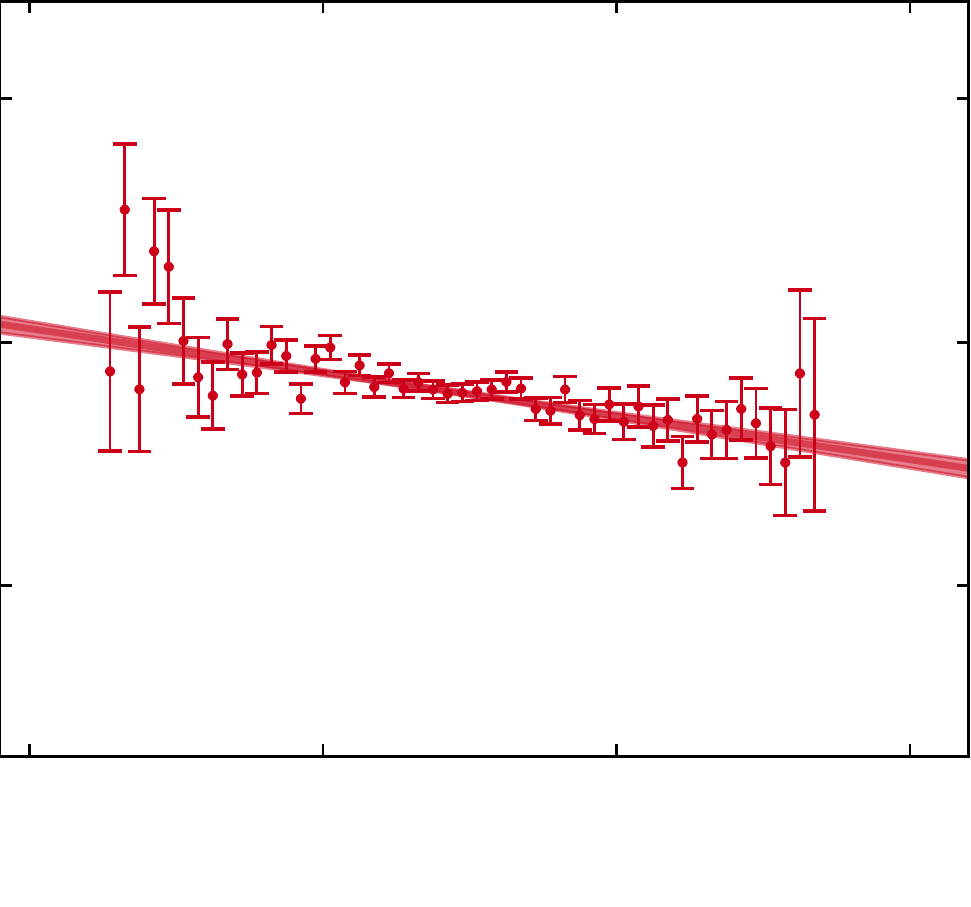}
\end{subfigure} \\

\begin{subfigure}{.6\columnwidth}
  \centering
  \includegraphics[width=\columnwidth]{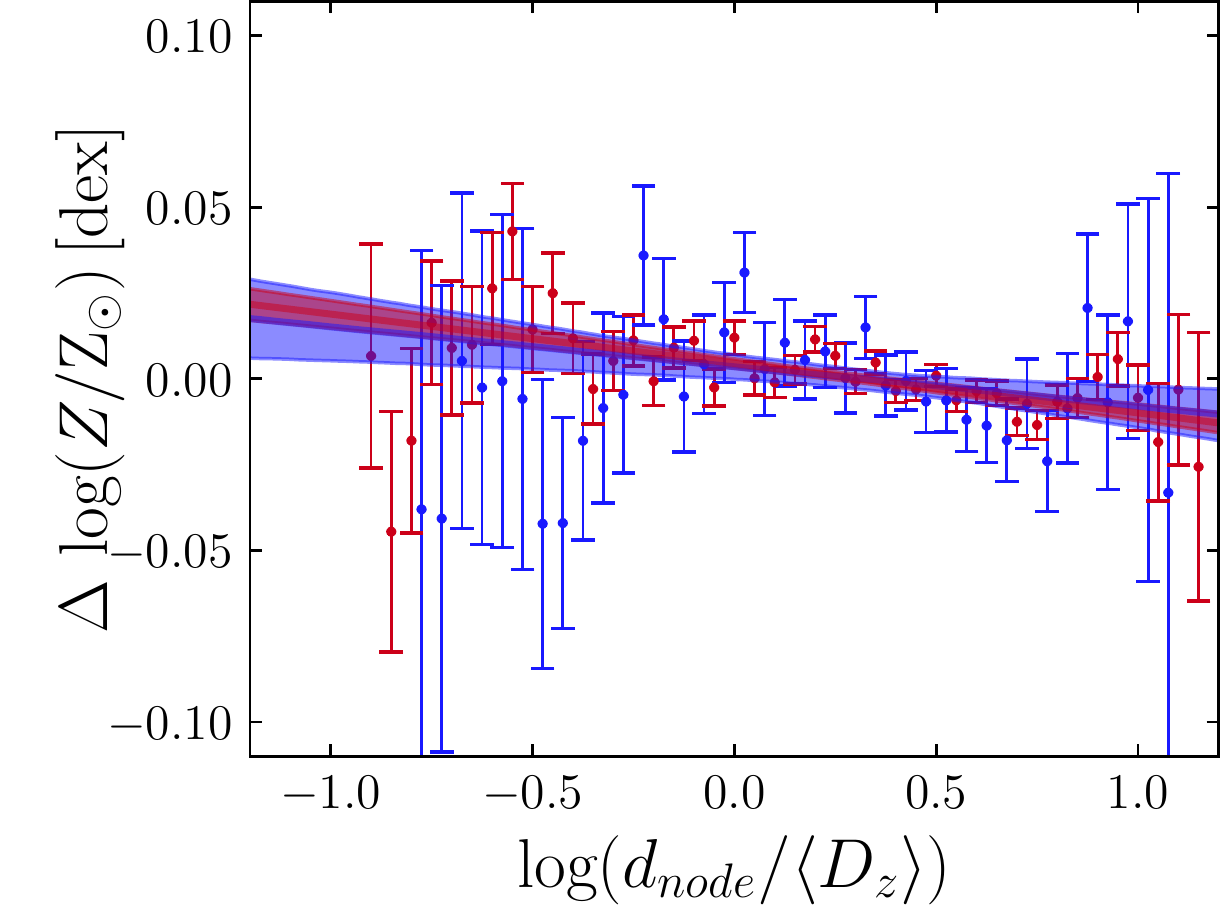}
\end{subfigure}%
\begin{subfigure}{.48\columnwidth}
  \centering
  \includegraphics[width=\columnwidth]{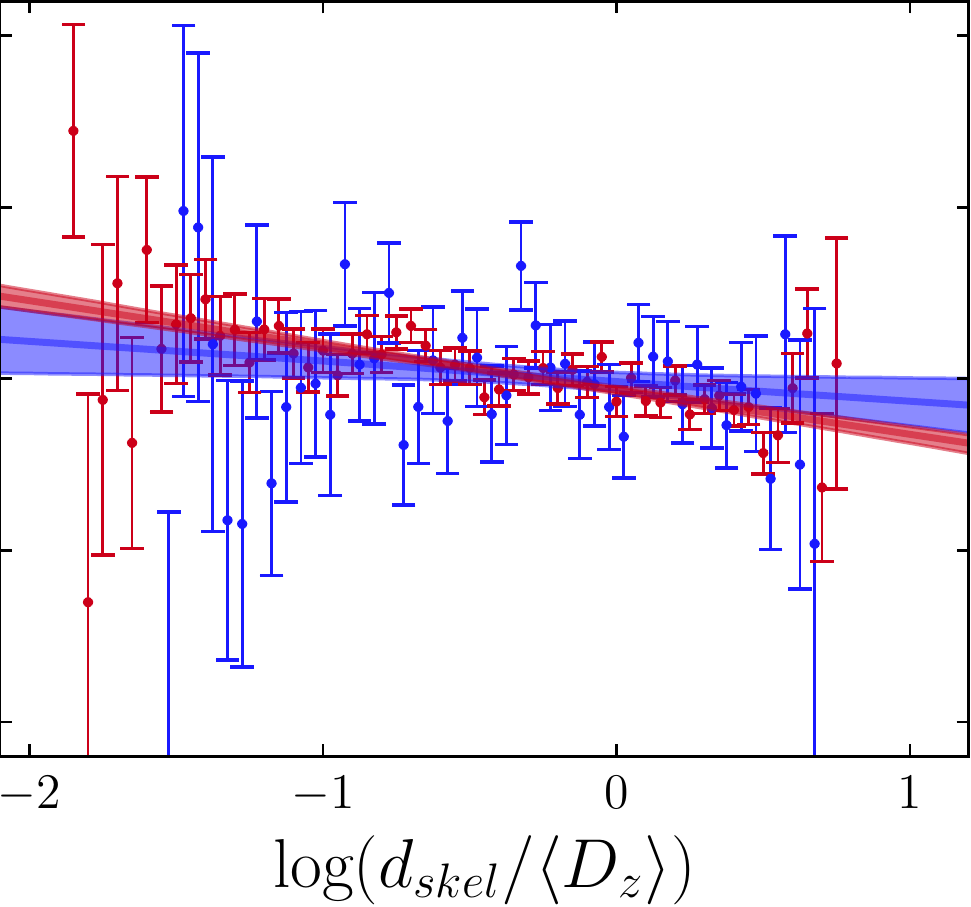}
\end{subfigure}%
\begin{subfigure}{.48\columnwidth}
  \centering
  \includegraphics[width=\columnwidth]{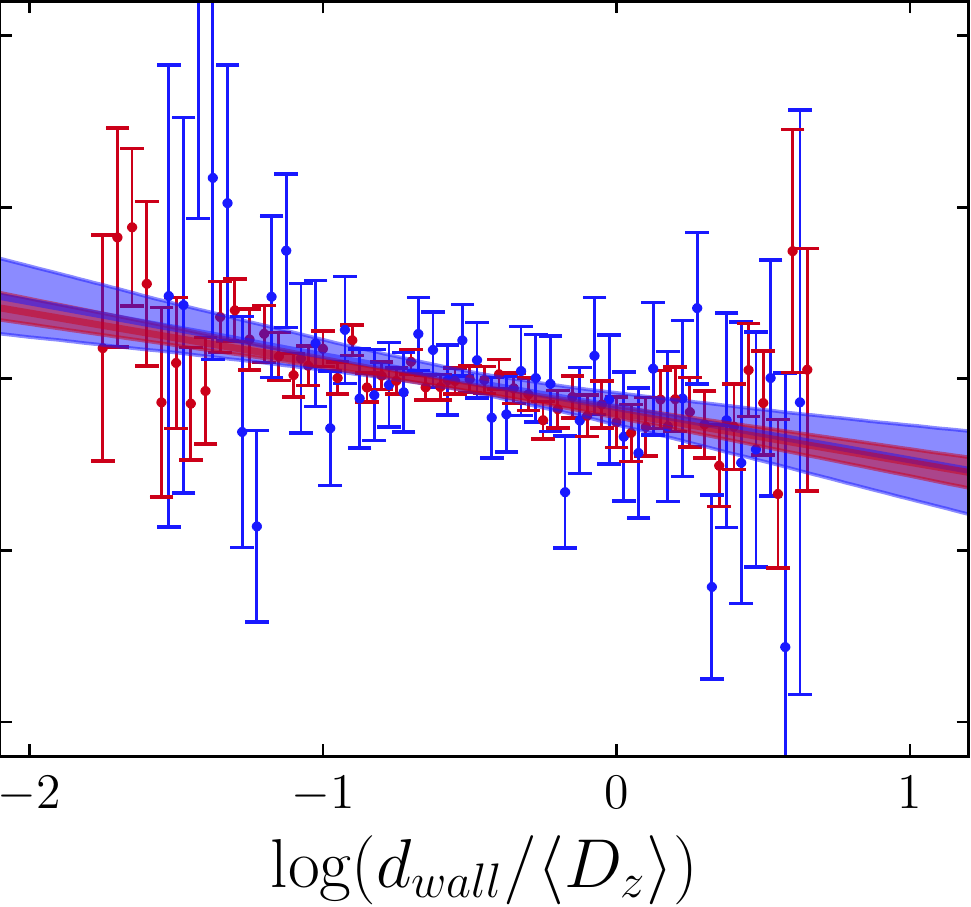}
\end{subfigure} \\

\caption{\label{fig:Z}Same as Fig.~\ref{fig:Age} but now for the weighted mean stellar metallicities of sample central galaxies. From left to right the panels show the gradients in distance to the cosmic web nodes, filaments and walls respectively. The bottom row shows the average offsets from the $Z$-$M_\star$ relations that were computed as described in Section~\ref{stn:Analysis of the Galaxy Property Offsets} of the appendix. Colour indicates halo mass and errorbars indicate the error of the mean. The thick continuous lines show the best fit linear model with the $1\sigma$ confidence region plotted as shaded areas. The slopes of the gradients can be found in Table~\ref{tab:Gradients_Z}. Close to each of the cosmic web features, central galaxies tend to be metal enhanced compared to their equal mass counterparts at large distances.}
\end{figure*}

\begin{table*}
 	\centering
	\caption{Same as Table~\ref{tab:Age} but now for stellar metallicity.}
	\label{tab:Z}
	\begin{tabular}{lcccccr} 
        \hline
        \multicolumn{1}{c|}{} & & \multicolumn{1}{c}{$k = node$}
                    & \multicolumn{1}{c}{$skel$}  & \multicolumn{1}{c}{$wall$}\\
                                    
        \hline
        \multirow{1}{*}{$\nabla_{k} ^{\Delta Z} \, [10^{-1} \textrm{dex/dex}]$}
     & field   
        & $-0.12 \pm 0.08$
        & $-0.05 \pm 0.05$
        & $-0.15 \pm 0.09$
        \\
     & groups  
        & $-0.14 \pm 0.03$	
        & $-0.13 \pm 0.01$
        & $-0.14 \pm 0.02$
        \\
        \\
         \multirow{2}{*}{$\textrm{log}(d_{0,k}^{\Delta Z} / \langle D_z \rangle) $ }
    &field    
        & $0.29 \pm 0.18$
        & -
        & $-0.54 \pm 0.33$
        \\
    & groups 
        & $0.31 \pm 0.07$	
        & $-0.25 \pm 0.03$
        & $-0.66 \pm 0.09$
        \\
        \\
        \multirow{2}{*}{$\delta_{k}^{Z}  \, [\textrm{dex}]$ }
     & field   
        & $0.022 \pm 0.015$
        & $-$
        & $0.032 \pm 0.020$
        \\
     & groups  
        & $0.029 \pm 0.008$	
        & $0.033 \pm 0.005$
        & $0.034 \pm 0.006$
        \\
        \\
        \multirow{1}{*}{$\delta_{M_h}^{Z} \, [\textrm{dex}]$}
    &
        & $0.064 \pm 0.005$
        & $0.059 \pm 0.005$
        & $0.056 \pm 0.005$
        \\
		\hline
	\end{tabular}
\end{table*}

While the luminosity weighted stellar ages are particularly sensitive to recent star formation \citep{Gallazzi2005}, the stellar metallicity $Z$ gives an estimate of the cumulative metal-enrichment of the stellar generations over the lifetime of a galaxy.
The first upper panels of Fig.~\ref{fig:Z} show the average stellar metallicities of low-mass central galaxies which only vary significantly with $d_{node}$ (at a $2.4\sigma$ level). At large distance to the node, low-mass centrals tend to be metal poor, while their counterparts in the vicinity of the nodes are metal rich. In the intermediate stellar mass bin, this increase of the average metallicity with decreasing $d_k$ is significant for each of the cosmic web features $k$ and for centrals in each halo mass bin (c.f. Table~\ref{tab:Gradients_Z}). As for the stellar ages, we attribute the constant metallicity offset between the two halo mass regimes of $\delta _{M_h}^Z \sim 0.06 \, \textrm{dex}$ to the imprint of the local environment (i.e. the central's halo mass) which is in qualitative agreement to the results presented in \citet{Pasquali2010} and \citet{Gallazzi2021}. In the most massive $M_\star - M_h$ bin, centrals also exhibit clear metallicity gradients $\nabla _k ^Z$ which do not depend systematically on the type of cosmic web feature $k$ (see Table~\ref{tab:Gradients_Z}). \par\noindent
The bottom row of Fig.~\ref{fig:Z} shows that the average metallicity offsets at given stellar mass $\Delta \textrm{log}(Z/\textrm{Z}_\odot )$ decrease significantly with increasing $d_{node}$, $d_{skel}$ and $d_{wall}$ for centrals in both halo mass regimes.
Towards the filaments, only the average metallicities of centrals in massive haloes (`groups' in Table~\ref{tab:Z}) are significantly affected. Here, the variation of the metallicity offset between far and close galaxies $\delta_{skel}^{Z}$ is significantly smaller (at a $3.6 \sigma$ level) than $\delta_{M_h}^{Z}$ which also applies to $\delta_{wall}^{Z}$ (at a $2.8\sigma$ level). \par\noindent
The transition distances $\textrm{log}(d_{0,k}^{Z} /\langle D_z \rangle)$ listed in Table~\ref{tab:Z} match well those found for the average stellar age offsets. Centrals turn from metal poorer to metal richer than average at a larger distance from the nodes than from the filaments and walls, respectively. Although the overall trend of decreasing $\Delta \textrm{log}(Z/\textrm{Z}_\odot )$ with increasing $d_{k}$ is similar, the different transition distances suggest a dissimilar role of the cosmic web features.

\subsection{\texorpdfstring{[$\alpha$/Fe]}{aFe}}
\label{stn:afe}

\begin{figure*}

\centering
\begin{subfigure}{.6\columnwidth}
  \centering
  \includegraphics[width=\columnwidth]{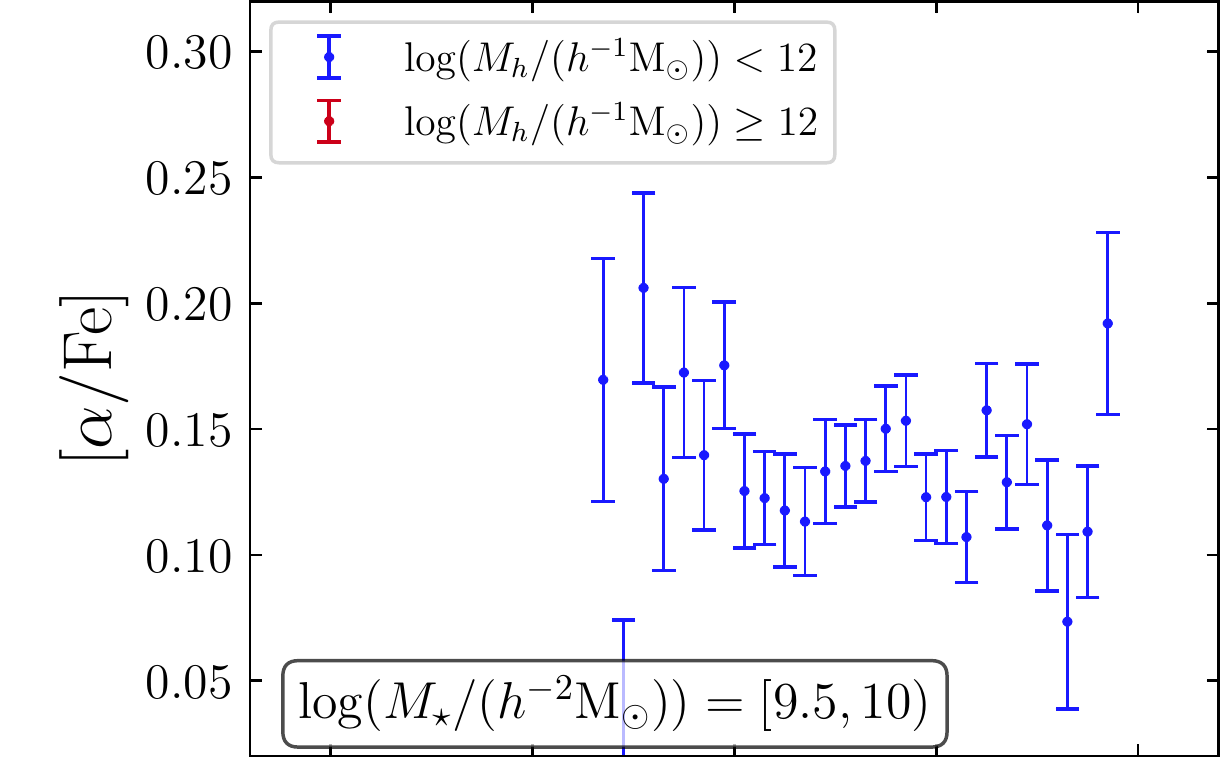}
\end{subfigure}%
\begin{subfigure}{.48\columnwidth}
  \centering
  \includegraphics[width=\columnwidth]{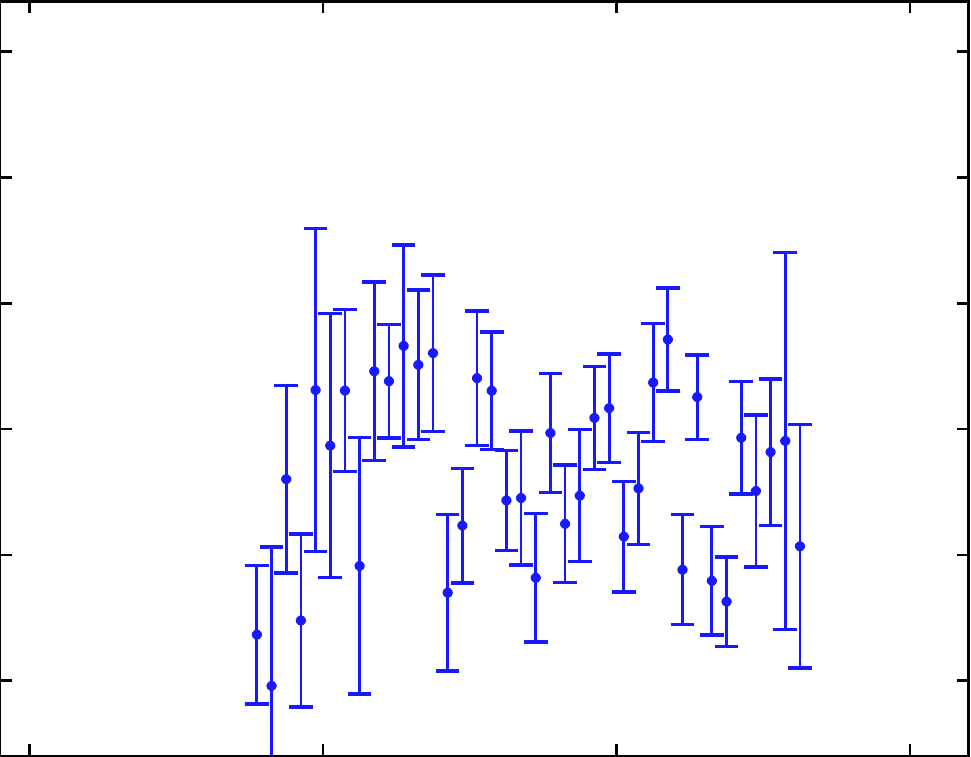}
\end{subfigure}%
\begin{subfigure}{.48\columnwidth}
  \centering
  \includegraphics[width=\columnwidth]{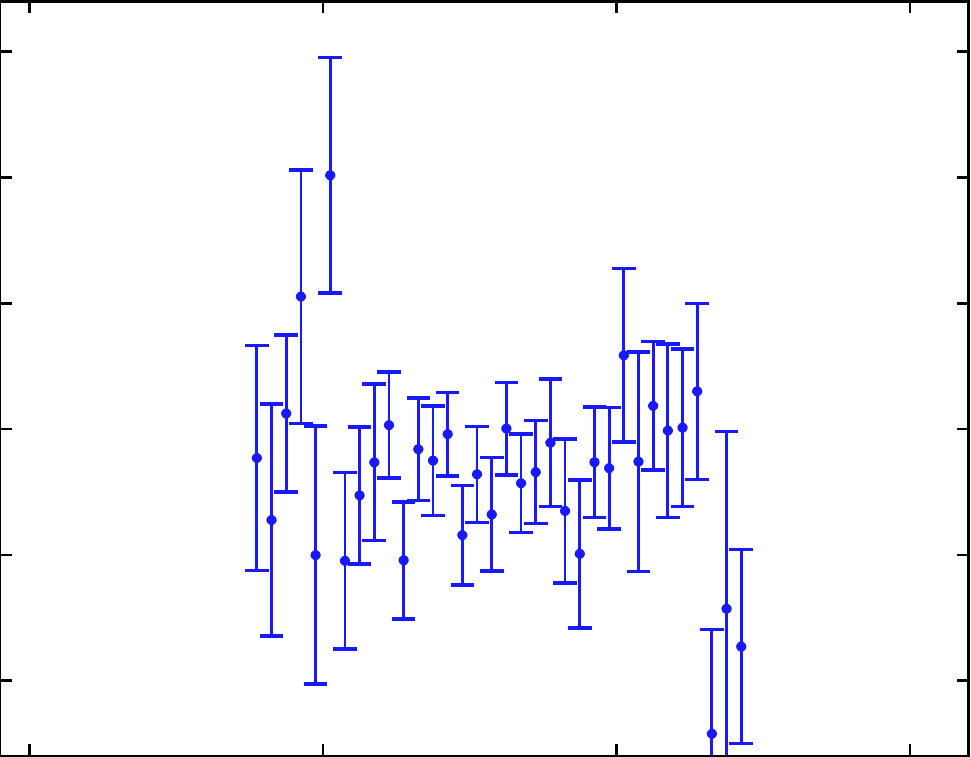}
\end{subfigure} 

\vskip -1pt

\begin{subfigure}{.6\columnwidth}
  \centering
  \includegraphics[width=\columnwidth]{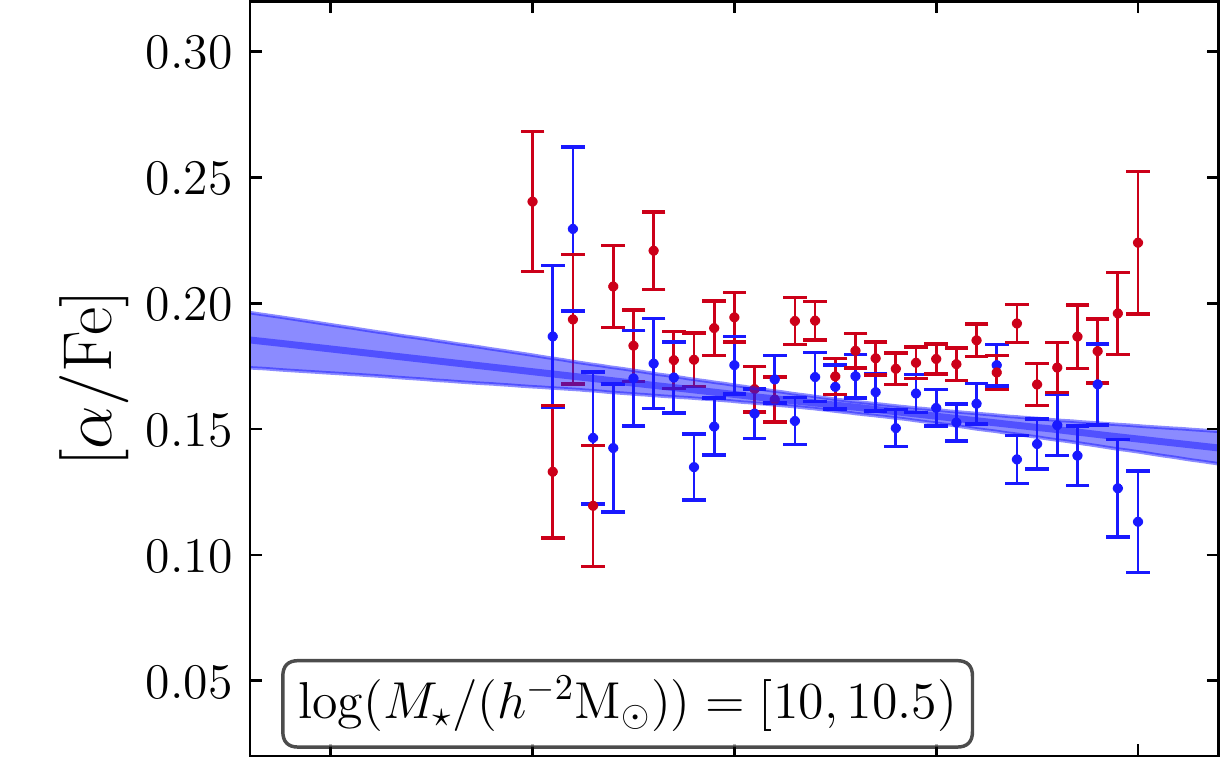}
\end{subfigure}%
\begin{subfigure}{.48\columnwidth}
  \centering
  \includegraphics[width=\columnwidth]{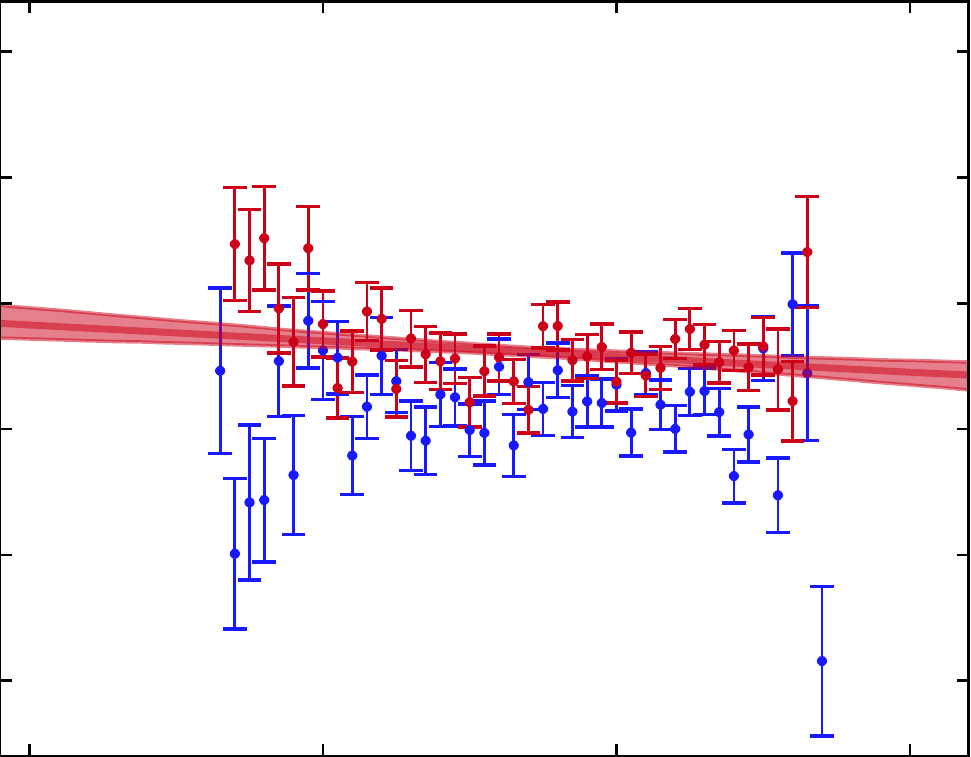}
\end{subfigure}%
\begin{subfigure}{.48\columnwidth}
  \centering
  \includegraphics[width=\columnwidth]{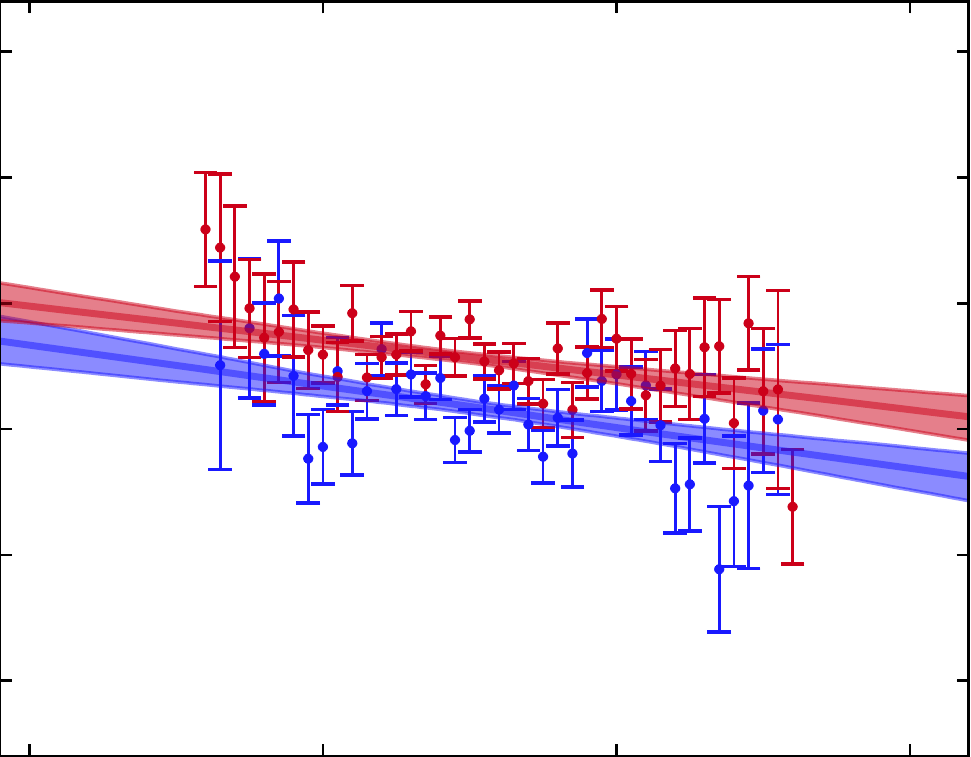}
\end{subfigure} 

\vskip -1pt

\begin{subfigure}{.6\columnwidth}
  \centering
  \includegraphics[width=\columnwidth]{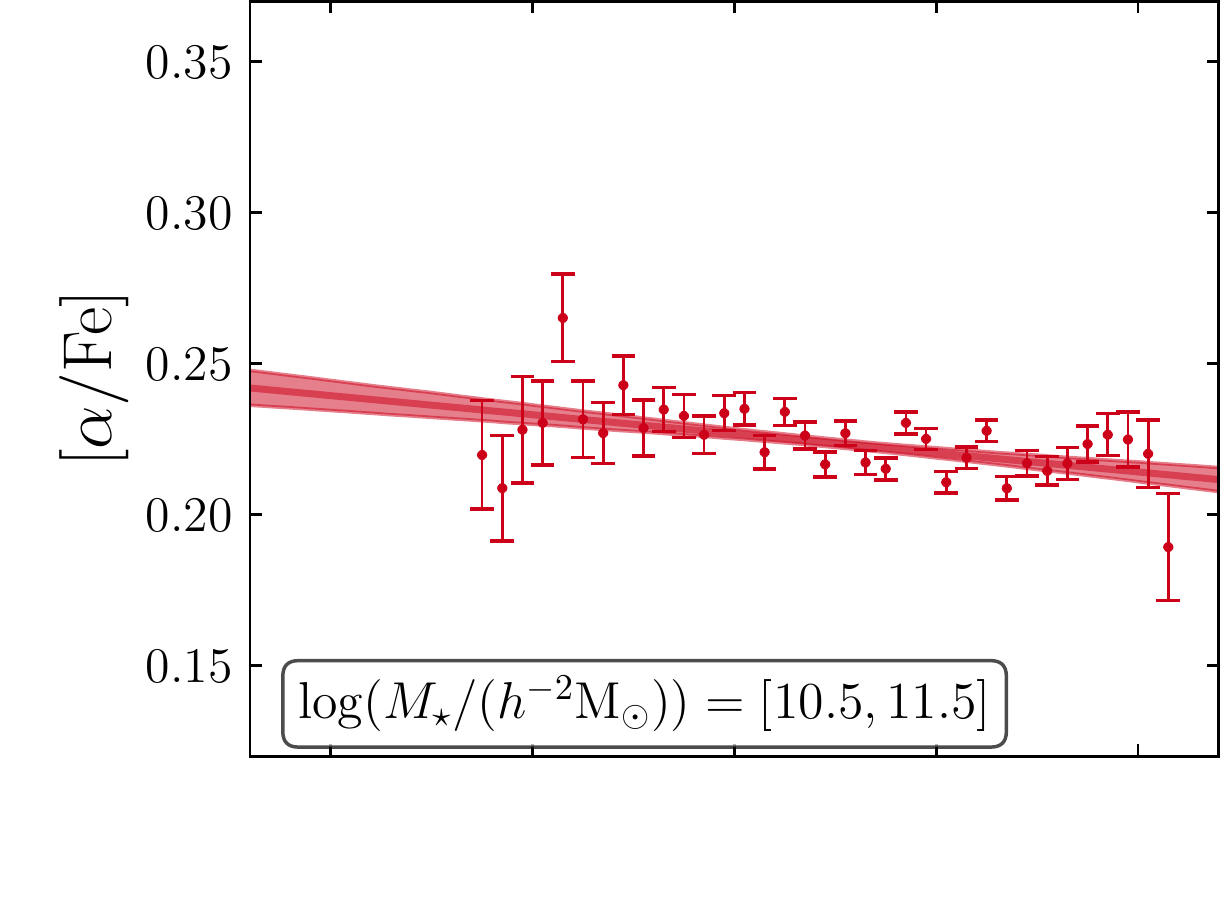}
\end{subfigure}%
\begin{subfigure}{.48\columnwidth}
  \centering
  \includegraphics[width=\columnwidth]{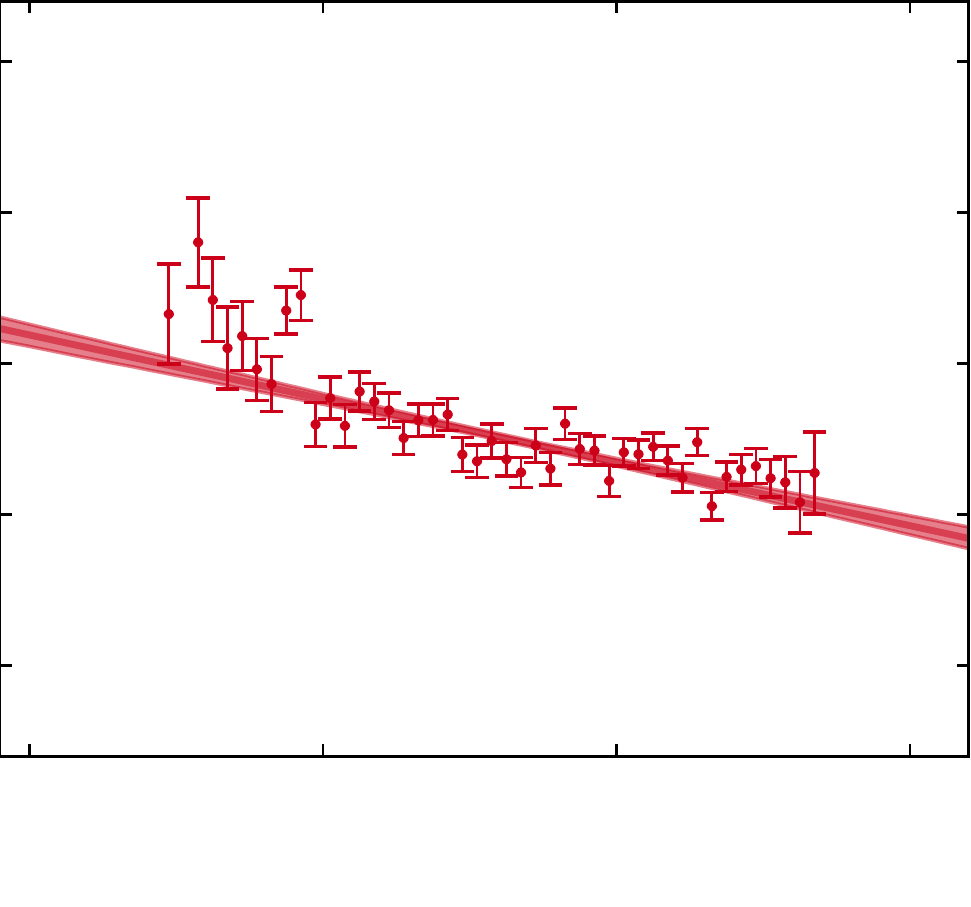}
\end{subfigure}%
\begin{subfigure}{.48\columnwidth}
  \centering
  \includegraphics[width=\columnwidth]{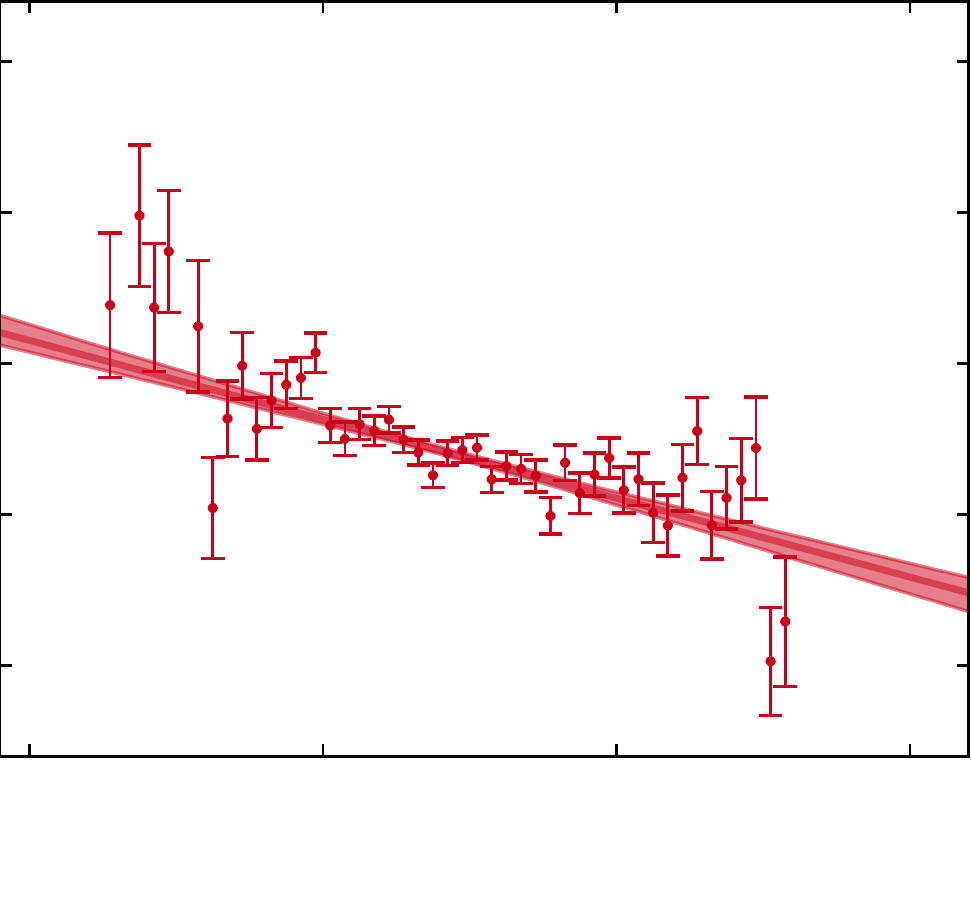}
\end{subfigure} \\

\begin{subfigure}{.6\columnwidth}
  \centering
  \includegraphics[width=\columnwidth]{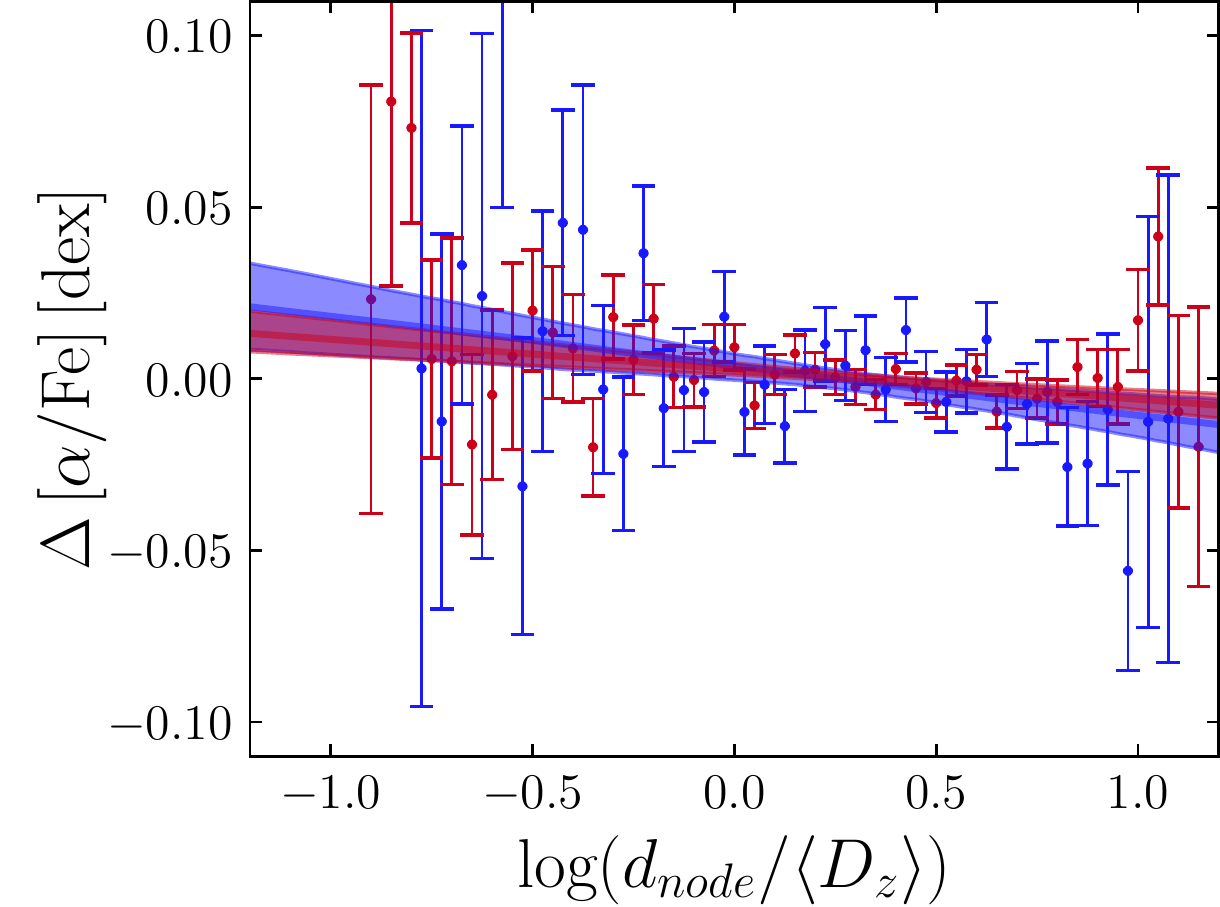}
\end{subfigure}%
\begin{subfigure}{.48\columnwidth}
  \centering
  \includegraphics[width=\columnwidth]{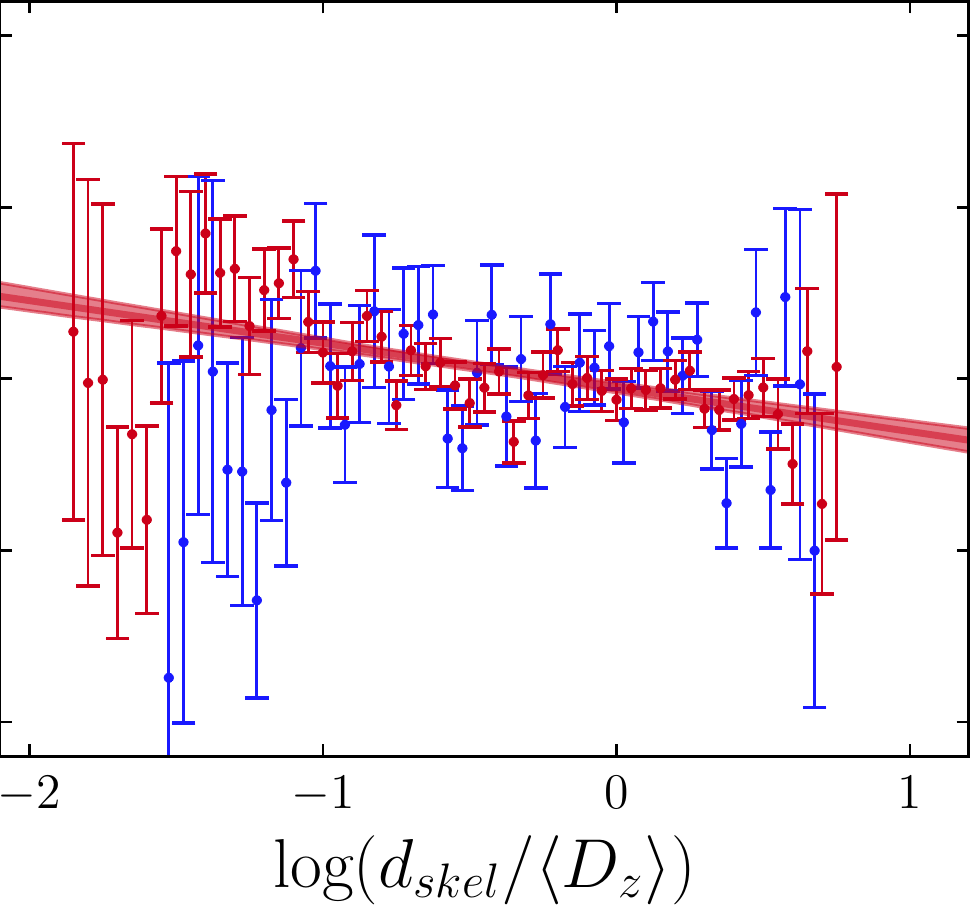}%
\end{subfigure}%
\begin{subfigure}{.48\columnwidth}
  \centering
  \includegraphics[width=\columnwidth]{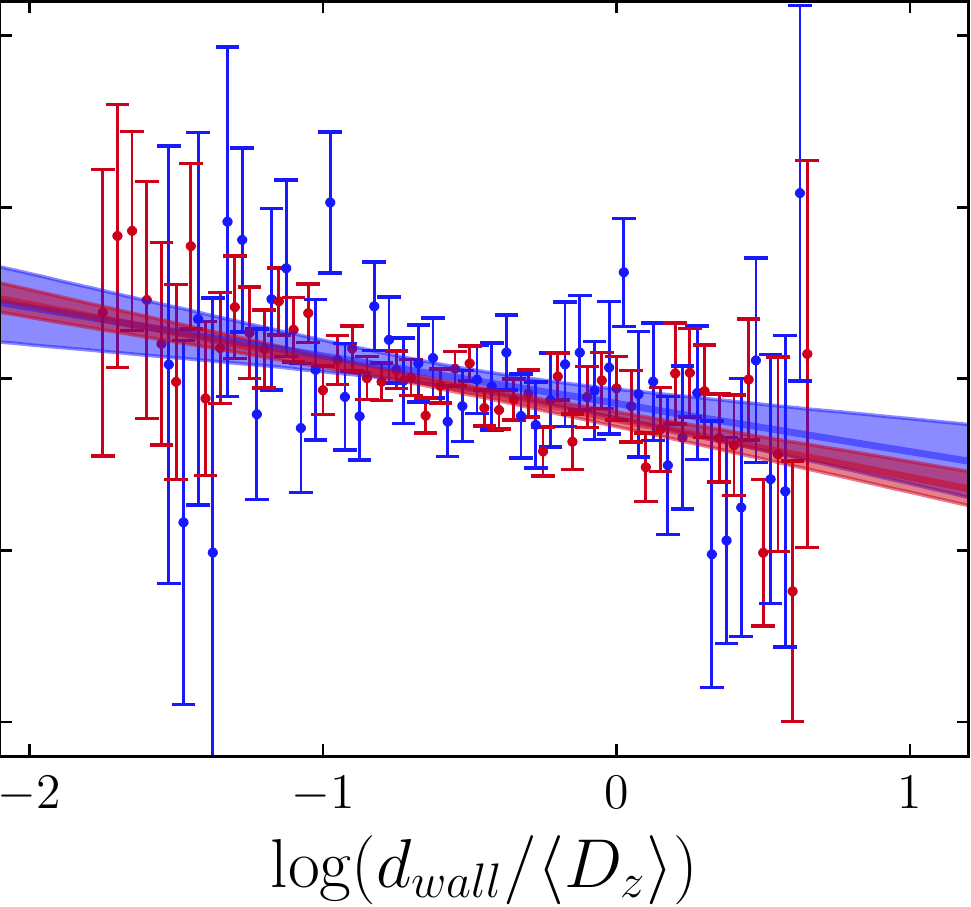}
\end{subfigure} \\

\caption{\label{fig:afe}Same as Fig.~\ref{fig:Age} for the average [$\alpha /$Fe] of the sample centrals. The corresponding slopes of the gradients are listed in the Table~\ref{tab:Gradients_afe} of the appendix. While lower mass centrals in less massive haloes are not significantly affected by the distance to the cosmic web features, those in more massive haloes exhibit decreasing [$\alpha /$Fe] with increasing $d_k$. This trend is strongest for the most massive central galaxies.}
\end{figure*}

\begin{table*}
 	\centering
	\caption{Same as Table~\ref{tab:Age} but now for [$\alpha$/Fe].}
	\label{tab:aFe}
	\begin{tabular}{lcccccr} 
        \hline
        \multicolumn{1}{c|}{} & & \multicolumn{1}{c}{$k = node$}
                    & \multicolumn{1}{c}{$skel$}  & \multicolumn{1}{c}{$wall$}\\
                                    
        \hline
        \multirow{1}{*}{$\nabla_{k} ^{\Delta [\alpha / \textrm{Fe}]} \, [10^{-1} \textrm{dex/dex}]$}
     & field   
        & $-0.14 \pm 0.08$
        & $-0.01 \pm 0.06$
        & $-0.14 \pm 0.07$
        \\
     & groups  
        & $-0.09 \pm 0.04$	
        & $-0.13 \pm 0.02$
        & $-0.18 \pm 0.04$
        \\
        \\
         \multirow{2}{*}{$\textrm{log}(d_{0,k}^{\Delta [\alpha / \textrm{Fe}]} / \langle D_z \rangle) $ }
    &field    
        & $0.27 \pm 0.16$
        & -
        & $-0.48 \pm 0.24$
        \\
    & groups 
        & $0.32 \pm 0.14$	
        & $-0.23 \pm 0.04$
        & $-0.69 \pm 0.15$
        \\
        \\
        \multirow{2}{*}{$\delta_{k}^{[\alpha / \textrm{Fe}]}  \, [\textrm{dex}]$ }
     & field   & $0.025 \pm 0.015$
        & $0.003 \pm 0.013$
        & $0.029 \pm 0.016$
        \\
     & groups  & $0.018 \pm 0.008$	
        & $0.034 \pm 0.007$
        & $0.039 \pm 0.011$
        \\
        \\
        \multirow{1}{*}{$\delta_{M_h}^{[\alpha / \textrm{Fe}]} \, [\textrm{dex}]$}
     &  
        & $0.019 \pm 0.002$
        & $0.025 \pm 0.002$
        & $0.017 \pm 0.002$
        \\
		\hline
	\end{tabular}
\end{table*}

As abundance ratio between elements that are primarily produced by two different mechanisms, [$\alpha$/Fe] is usually considered to be a proxy for the duration of a galaxy's star formation episode: $\alpha$ elements are predominantly synthesised in the $\alpha$-process during the secular evolution of a massive star, and then released into a galaxy's ISM when the star explodes as a supernova type II. On the other hand, supernovae type Ia (SNIa) are thought to be predominantly responsible for the Fe-enrichment of the ISM which happens over several Gyrs.
Therefore, galaxies which experienced a short but intense star-burst phase (like massive early-type galaxies) exhibit a high [$\alpha$/Fe], while galaxies undergoing a more prolonged star formation activity have lower values of [$\alpha$/Fe].

Fig.~\ref{fig:afe} shows the dependence of the average [$\alpha$/Fe] abundance ratio on distance to the cosmic web nodes, filaments and walls. Only above $M_\star = 10^{10}\,h^{-2}\textrm{M}_\odot$ consistent gradients $\nabla _k ^{[\alpha / \textrm{Fe}]}$ start to emerge: Massive centrals exhibit mildly decreasing average [$\alpha /$Fe] with increasing distance to each of the features $k$, pointing to more extended star formation histories for massive centrals at large distance to the cosmic web. For the nodes (filaments) this trend is only significant for centrals in the low (high) halo mass regime at a $ 1.7\sigma$ ($2.3\sigma$) level (see Table~\ref{tab:Gradients_afe}).
In the most massive $M_\star$-$M_h$ bin, the average [$\alpha$/Fe] of central galaxies is sensitive to each of the cosmic web distances $d_k$. In contrast to what is observed for the average stellar ages and metallicities, the trends in [$\alpha$/Fe], as quantified by the gradients $\nabla _k ^{[\alpha / \textrm{Fe}]}$ become more evident with increasing stellar mass. \\
The average offsets from the [$\alpha$/Fe]-$M_\star$ scaling relation in the bottom row of Fig.~\ref{fig:afe} show that centrals with $M_h \geq 10^{12} \, h^{-1}\textrm{M}_\odot$ exhibit higher-than-average [$\alpha$/Fe] in the proximity of each cosmic web feature $k$, which decreases with increasing $d_{k}$.
On the other hand, centrals in the low halo mass regime are only affected significantly by the nodes and walls. We notice that the variation in [$\alpha$/Fe] produced by the cosmic web $\delta_{k}^{[\alpha/\textrm{Fe}]}$ is consistent within 2$\sigma$, with the change $\delta_{M_h}^{[\alpha/\textrm{Fe}]}$ produced by the halo mass of the central (see Table~\ref{tab:aFe}). 
The transition distance at which the offsets $\Delta [\alpha/\textrm{Fe}]$ turn from positive to negative decreases from nodes to filaments and walls, and is comparable with what estimated for stellar ages and metallicities.

\section{Discussion}
\label{stn:Discussion}


Contemporary models of galaxy evolution predict a hierarchical cosmic structure growth where dark matter haloes grow gradually in a bottom up fashion. In this picture, the local environment can be described by the properties of the central's extended dark matter halo. Beyond its virial radius, the anisotropic skeleton of the large scale structure forms a framework in which the individual dark matter haloes are embedded. Within the metric defined in Section~\ref{stn:The Cosmic Web Metric}, cosmic web environment is parametrised by the distance $d_k$ to the cosmic web features.  In our analysis we focus on central galaxies exclusively, since they are less affected by their host halo than satellites.


We observe that for both halo mass regimes (i.e. field vs. group environment) central galaxies at small distance to the cosmic web features exhibit on average {\it i)} lower specific star formation rates, {\it ii)} higher stellar metallicity, and {\it iii)} slightly higher [$\alpha$/Fe] abundance ratios than their equal mass counterparts at large distances. These trends persist even when we control for galaxy stellar mass by computing the average property offsets from the property - $M_\star$ relation. Thus, the gradients that we detect in the average property offsets are driven by the large-scale environment, and are not a mere reflection of the galaxies' scaling relations and their mass segregation towards the cosmic web features. These observed trends indicate that not only the nodes of the cosmic web (its higher density peaks) but also its filaments and walls are able to quench by some degree the star formation activity of central galaxies independently of their local environment (parametrised here by the halo mass).\\
Between centrals in a field or group environment we do not identify a systematic difference. We recall that our binning in halo mass primarily separates isolated field centrals from group/cluster centrals (see Section~\ref{stn:The Working Sample}) where the latter are dominated by centrals in haloes with $M_h < 10^{13}\,h^{-1}\textrm{M}_\odot$. Therefore, we do not properly distinguish galaxy groups from clusters which range up to $\sim 10^{15}\,h^{-1}\textrm{M}_\odot$. We can state, however, that centrals in small groups and isolated centrals are affected by the cosmic web features in a similar fashion.\\
The gradients in the average property offsets may be used to estimate the spatial extent over which the different cosmic web features are able to affect galaxy properties. There exists indeed a transition distance where these average property offsets turn from positive to negative which for each cosmic web feature, {\it i)} does not depend on the galaxy property analyzed, and {\it ii)} decreases as we move from nodes to filaments and walls {\it iii)} does not depend on the halo mass of the central.
Assuming $\langle D_z \rangle = 5\,\textrm{Mpc}$ (the mean-galaxy separation of our sample at $z = 0$), the median transition distances are $d_{0,node} = 9.8\, \textrm{Mpc} $, $d_{0,skel} = 2.2\,\textrm{Mpc}$ and $d_{0,wall} = 1.2\, \textrm{Mpc}$.
This hierarchy is expected by construction of the metric. While the nodes sample the rare high density peaks of the cosmic web, filaments and especially walls occur more frequently, hence occupy a larger fraction of the volume. Despite this fundamental difference, the property gradients have similar slope towards each of the cosmic web features. This suggests that the cosmic web nodes, filaments and walls have a similarly strong influence on the evolution of central galaxies, but act on different spatial scales. \\


Our results are qualitatively consistent with the findings from cosmological simulations. For instance, \cite{Xu2020} showed that in the EAGLE simulation \citep{Schaye2015} central galaxies with $M_\star < 1.8 \times 10^{10} \, h^{-1}\textrm{M}_\odot$ have redder colors, lower sSFRs and higher metallicities in nodes than those in filaments, sheets and voids. 
Using IllustrisTNG-100, \cite{Martizzi2020} found that at given halo mass, galaxies with stellar masses lower than the median value are more likely to be found in voids and sheets, whereas galaxies with stellar masses higher than the median are more likely to be found in filaments and knots. The projected gas density and temperature are seen to be three orders of magnitude higher in nodes and filaments than in under-dense voids \citep{Martizzi2019}. The hot gas stretches out for several Mpc around nodes, and is less extended around filaments and walls. In addition, the gas moves at higher velocity when it is advected onto filaments and nodes, while in their centres it moves slower. A recent study from \cite{Galarraga-Espinosa2020} showed that in in TNG-300 the filament cores out to \mbox{1.5 Mpc} from the spines are dominated by hot and dense gas, with average temperatures of $4-13 \times 10^5 \, \textrm{K}$. \par\noindent
We thus postulate that, in these conditions, the hot gas in the cosmic web is capable of removing the centrals' cold gas reservoir via ram pressure stripping \citep{Gunn1972}, qualitatively similar to how the intra-group or intra-cluster medium acts on the satellite population of the centrals' host haloes (\citealt{Boselli2006, Pasquali2015}). The occurrence of such {\it cosmic web stripping} has already been suggested by \citet{Benitez-Llambay2013} in order to explain the small number of dwarf galaxies compared to the numerous low-mass haloes predicted by $\Lambda$CDM. The increasing matter density towards the nodes, filaments and walls of the cosmic web could deprive central galaxies of their hot gas reservoir via {\it strangulation} \citep{Larson1980}, in a similar qualitative fashion to how the centrals' host haloes affect their satellites (\citealt{Boselli2006, Pasquali2015}). \par\noindent
Our analysis shows that central galaxies tend to be metal richer than average in proximity of nodes, filaments and walls, and metal poorer at large distances. We expect the cosmic web stripping to remove the metal-poor gas in the galaxy outskirts, preventing it from flowing into the galaxy central regions and diluting the metallicity of the ISM there in, qualitatively similar to what \citet{Bahe2017} find occurring in galaxy clusters in the EAGLE simulation. Following the argumentation of \citet{Pasquali2012, Pasquali2019}, the metal-rich gas in the galaxy central regions (those sampled by the SDSS fiber) continues forming stars that are rich in metals and increases the metallicity of the ISM throughout their evolution \citep[see also][]{Maier2019,Trussler2020, Gallazzi2021}. With the gas density increasing towards the cosmic web features we expect the cosmic web stripping to become more efficient, and hence to foster a higher stellar metallicity in those galaxies closer to nodes, filaments and walls. \par\noindent
As a net effect of these processes the star formation activity of the centrals closer to nodes, filaments and walls is reduced, while allowing those centrals farther away to grow by smooth gas accretion \citep{Dubois2012,Welker2017}. \par\noindent
Still, there must be some difference between the action of cosmic web on central galaxies and the action of group/cluster environment on satellite galaxies. In fact, at fixed stellar mass centrals are observed to be on average younger and more star-forming that satellites \citep{vandenBosch2008, Pasquali2009, Wetzel2012}. \\


Another viable explanation for the observed trends is that galaxies may have formed differently depending on their position in the cosmic web. As reported by \cite{Kraljic2018}, galaxies in the proximity of any cosmic web feature tend to be more massive on average. Further, it is well-established that massive galaxies consume their available gas more 
efficiently and quickly, hence they build up their stellar mass on shorter time scales, an effect known as $downsizing$ \citep{Cowie1996, Fontanot2009}. At the same time, observations \citep{Weinberg1997, Rauch1997,CenOstriker2006, Dave2001, Cantalupo2014, LeeWhite2016, Umehata2019} and simulations \citep{Shull2012, Nevalainen2015, Snedden2016, Martizzi2019, Galarraga-Espinosa2020} indicate that the cosmic web contains large amounts of gas at both low and high redshifts. Therefore it is reasonable to expect that galaxies forming near the cosmic web features are naturally prone to grow more massive, older and metal-richer since their efficient star formation quenches them earlier and increases their metal content more than galaxies farther away from the cosmic web. Such star formation activity is expected to chemically pollute the gas in the cosmic web features via galactic outflows. When this gas is later accreted by the local galaxies, it contributes to further increase their stellar metallicity. \par\noindent
In addition, the higher galaxy density closer to the cosmic web likely fosters mergers which contribute not only to build up galaxy stellar mass (especially when they are major mergers), and stellar metallicity thanks to enhanced star formation (when they are gas rich, e.g. \citealt{DiMatteo2007}), but also to quench the remnants on relatively short time scales. The increasing fraction of ellipticals with respect to spiral galaxies with decreasing distance to the cosmic web filaments reported by \cite{Kuutma2017} can be interpreted as a morphological transformation induced by galaxy mergers (see also \citealt{Codis2012, Dubois2014}). Therefore, a higher occurrence of galaxy mergers in the vicinity of the cosmic web is a complementary explanation of why we observe older, less star forming, metal richer and slightly $\alpha$-enhanced central galaxies at smaller distances from nodes, filaments and walls.\newline
Hydrodynamic simulations suggest that smooth accretion of cold gas can provide dark matter haloes with pristine gas, especially in the filament outskirts \citep{Keres2005, DekelBirnboim2006, Pichon2011, Cautun2014}. This gas can feed a prolonged star formation activity in their central galaxies \citep[e.g.][]{Martin2016}, still keeping the galaxies stellar metallicity lower than what observed near the cosmic web features. This accretion can thus account for the larger star formation rates, younger stellar ages and lower stellar metallicity that we observe for central 
galaxies away from the cosmic web features. 


Independent of the nature or nurture origin of the observed property gradients, the cosmic web quenching can easily explain why we see central galaxies forming stars at lower rate and having older stellar populations in the proximity of the cosmic web features. It may also account for the non-negligible fraction of passive galaxies observed beyond the virial radius of galaxy clusters \citep[e.g.][]{Wetzel2012}.
Once those pre-processed galaxies have been accreted onto their present-day halo they are recognised as satellite galaxies. 
From the mass assembly history of present-day satellites, \cite{Smith2019} found that ancient infallers ( >5 Gyr ago) are on average characterised by a reduced mass growth at lookback times as early as \mbox{9 Gyr} (i.e. before they are expected to be accreted onto the progenitor of their present-day host halo). The authors argue this could be an effect of the large scale structure at those epochs, which is in line with our findings on how the cosmic web features affect the star formation activity at z $\sim$ 0.
Another possible signature of the cosmic web quenching can be found in the results of \citet{Gallazzi2021}, who register a small but systematic excess of few hundred Myr in the luminosity-weighted stellar age of quiescent satellites in massive host haloes with respect to quiescent centrals at z $\sim$ 0. Since these galaxies are on average as old as 8 Gyr, they likely had their star formation activity quenched by the cosmic web at those epochs and before falling into the progenitor of their present-day host haloes. It appears that processes induced by the cosmic web are required to obtain a complete explanation for the origin of the galaxy population present in groups and clusters. \par\noindent
Given that within groups and clusters low-mass satellites are more sensitive to environmental processes \citep{Bekki2009, Pasquali2010, Benitez-Llambay2013}, it would be interesting to specifically investigate the dependence of the observed gradients for low stellar mass (i.e. $< 10^{9.5} \, h^{-2}\textrm{M}_\odot$) and high halo mass (i.e. $> 10^{14} \, h^{-1}\textrm{M}_\odot$) galaxies. If `nurture' as exerted by the cosmic web played a substantial role in shaping the properties of galaxies before their accretion
onto a local host halo, we would expect that lower mass satellites in more massive haloes are most affected by it.  \\

\section{Summary}
\label{stn:Summary}

We used the SDSS DR7 group catalogue from \citet{Wang2014}, the stellar properties catalogue from \cite{Gallazzi2021} and the catalogue of distances to the cosmic web from \citet{Kraljic2020b} as computed with {\small DisPerSE} \citep{Sousbie2011} to investigate how the stellar properties of central galaxies depend on their distance to the 3D cosmic web features. In our metric the nodes, filaments and walls represent overdense regions of the cosmic web with 0-, 1- and 2- dimensional extent, respectively. 

In each of the analysed properties sSFR, stellar age, metallicity and element abundance ratio [$\alpha$/Fe]) central galaxies show significant gradients w.r.t. cosmic web environment. We qualitatively confirmed the sSFR gradients towards the cosmic web filaments that have already been identified for galaxies in the GAMA survey by \cite{Kraljic2018}, the VIPERS survey by \cite{Malavasi2017} and COSMOS by \cite{Laigle2018}. Further, we found that central galaxies at large distances to the cosmic web features have on average younger luminosity-weighted stellar ages and lower stellar metallicities compared to their equal mass counterparts at small cosmic web distances. Also the [$\alpha$/Fe] abundance ratio of massive centrals is higher than average if they are located in the proximity of the cosmic web features. \par\noindent 
We have shown that these gradients are not only due to mass segregation towards the cosmic web features, but remain when we compute the offsets in each observed galaxy property from the scaling relation between that property and galaxy stellar mass. Furthermore, the amplitudes of the property variations from nearby to distant central galaxies indicate that the imprint of the cosmic web features on stellar age, metallicity and [$\alpha$/Fe] is similarly strong as the difference observed between the two halo mass regimes.\par

A picture emerges, where central galaxies at large cosmic web distances assemble their stellar mass later and experience a prolonged star formation activity. Centrals do not experience the effects that environment applies to satellites, but they undergo mergers/interactions that make them grow
in mass, accretion of intra-cluster gas which is kept hot by the host and their AGN and hence can not be used to form new stars.
The centrals' higher-than-average present-day sSFR, young stellar ages and low metallicities at large distance from the cosmic web features are possibly a result of cold gas accretion from their surroundings. Close to the cosmic web features these properties deviate from average in the opposite manner, which could be a reflection of quenching processes acting on central galaxies similar to those that satellites experience in galaxy groups and clusters, i.e. strangulation, ram pressure stripping, galaxy interactions and mergers. 
Alternatively, the lower sSFR, older age and higher stellar metallicity of centrals close to the cosmic web could be owed to them forming in a gas- and metal-rich environment (where also mergers are more frequent), through efficient star formation which quenches them on relatively short timescales.\\

A similar analysis to ours for galaxies in the next generation redshift surveys (e.g. 4MOST, Euclid, HSC, PFS) could help to shed light on the role of the cosmic web at a time when galaxies assembled the bulk of their present day stellar mass.
Furthermore, cosmological hydrodynamical simulations could provide valuable insight on gas cold streams and their relationship to the properties of central galaxies, depending on their position in the cosmic web. 
This would help to constrain the nature vs. nurture origin of the observed trends, i.e. whether they are a reflection of the central galaxies' formation history or whether quenching processes exerted by the cosmic web are in charge.

\section*{Acknowledgements}
We thank the anonymous referee for suggestions and comments that helped improve the presentation of this work. 
We thank St\'ephane Rouberol for the smooth running of the HORIZON Cluster, where some of the post-processing was carried out.
We thank Thierry Sousbie for provision of the {\small DisPerSE} code (\href{http://ascl.net/1302.015}{ascl.net/1302.015}).
NW thanks Eva K. Grebel for logistic support at ARI where this project was carried out. NW and TMJ are fellows of the International Max Planck Research School for Astronomy and Cosmic Physics at the University of Heidelberg (IMPRS-HD). KK thanks Christophe Pichon for his exceptional support and enlightening discussions.

\section*{Data Availability}
The data underlying this article are available from the public sources in the references or links given in the article (or references therein).



\bibliographystyle{mnras}
\bibliography{MN-21-0777-MJ.R1_final} 




\appendix

\section{Gradients of the Galaxy Properties}
\label{stn:appendix}
In the following, we list the gradients of the average galaxy properties with cosmic web environment which were obtained as described in Section~\ref{stn:Methodology}.
For each average galaxy property $\bar{y}$ a separate table shows the gradients $\nabla _k ^{\bar{y}}$ w.r.t. the distance to the cosmic web features $k$ in each $M_\star - M_h$ bin. Furthermore, the gradients of the average galaxy property offsets $\nabla _k ^{\Delta \bar{y}}$ are shown.

\begin{table}
	\centering
	\caption{Distance gradients of the weighted average \mbox{sSFR$_{\textrm{glo}}$} towards the cosmic web features. For each feature $k$ we give the gradient $\nabla _k ^{\textrm{sSFR}_{\textrm{glo}}}$ in each $M_\star$ bin in the first three rows, and the average $\Delta \textrm{sSFR}_{\textrm{glo}}$ gradients in the fourth row (indicated by $\Delta$). The two halo mass regimes (i.e. field and group centrals) are split into two columns. The values of the gradients are given in units of dex/dex. 
    In $M_\star$-$M_h$ bins where the number of galaxies was not sufficient to robustly determine the slope of the property gradient via linear regression no values are shown.}
	\label{tab:Gradients_sSFR}
	\begin{tabular}[!htb]{*{9}{c}}

&\multicolumn{1}{c|}{$M_\star$} & \multicolumn{1}{c}{field}
            & \multicolumn{1}{c}{groups}\\
                            
\hhline{====}

\multirow{4}{*}{Nodes}
&\multicolumn{1}{c|}{[9.5,10)}
& $(0.70 \pm 0.58) \times 10^{-1}$	
& - 	
\\
&\multicolumn{1}{c|}{[10,10.5)}
& $(1.44 \pm 0.23) \times 10^{-1}$	
& $(1.82 \pm 0.35) \times 10^{-1}$		
\\
&\multicolumn{1}{c|}{[10.5,11.5]}
& - 	
& $(1.75 \pm 0.21) \times 10^{-1}$					
\\
\noalign{\smallskip}
&\multicolumn{1}{c|}{$\Delta$}
& $(0.71 \pm 0.21) \times 10^{-1}$	
& $(1.01 \pm 0.15) \times 10^{-1}$					
\\
\hline 

\multirow{4}{*}{Filaments}
&\multicolumn{1}{c|}{[9.5,10)}
& $(0.87 \pm 0.39) \times 10^{-1}$	
& - 		
\\
&\multicolumn{1}{c|}{[10,10.5)}
& $(1.16 \pm 0.20) \times 10^{-1}$	
& $(1.49 \pm 0.22) \times 10^{-1}$					\\
&\multicolumn{1}{c|}{[10.5,11.5]}
& - 	
& $(2.16 \pm 0.16) \times 10^{-1}$						
\\
\noalign{\smallskip}
&\multicolumn{1}{c|}{$\Delta$}
& $(0.56 \pm 0.17) \times 10^{-1}$	
& $(1.11 \pm 0.12) \times 10^{-1}$			
\\
\hline 

\multirow{4}{*}{Walls}
&\multicolumn{1}{c|}{[9.5,10)}
& $(0.59 \pm 0.51) \times 10^{-1}$	
& - 			 \\
&\multicolumn{1}{c|}{[10,10.5)}
& $(0.64 \pm 0.21) \times 10^{-1}$	
& $(1.47 \pm 0.24) \times 10^{-1}$	
\\
&\multicolumn{1}{c|}{[10.5,11.5]}
& - 	
& $(1.44 \pm 0.25) \times 10^{-1}$			
\\
\noalign{\smallskip}
&\multicolumn{1}{c|}{$\Delta$}
& $(0.37 \pm 0.22) \times 10^{-1}$	
& $(1.05 \pm 0.15) \times 10^{-1}$					
\\

\hline
\end{tabular}
\end{table}

\begin{table}
	\centering
	\caption{Same as Table~\ref{tab:Gradients_sSFR} but for the gradients of the average stellar age $\nabla _k ^{age}$ and the gradients of the average offsets at given stellar mass $\nabla _k ^{\Delta age}$. The values are given in units of dex/dex.}
	\label{tab:Gradients_Age}
	\begin{tabular}[!htb]{*{9}{c}}

&\multicolumn{1}{c|}{$M_\star$} & \multicolumn{1}{c}{field}
            & \multicolumn{1}{c}{groups}\\
                            
\hhline{====}

\multirow{4}{*}{Nodes}
&\multicolumn{1}{c|}{[9.5,10)}
& \textcolor{gray}{$(-0.21 \pm 0.27) \times 10^{-1}$}	
& - 				  \\
&\multicolumn{1}{c|}{[10,10.5)}
& $(-0.42 \pm 0.08) \times 10^{-1}$	
& $(-0.34 \pm 0.06) \times 10^{-1}$		
\\
&\multicolumn{1}{c|}{[10.5,11.5]}
& - 	
& $(-0.28 \pm 0.02) \times 10^{-1}$			
\\
\noalign{\smallskip}
&\multicolumn{1}{c|}{$\Delta$}
& $(-0.28 \pm 0.09) \times 10^{-1}$	
& $(-0.24 \pm 0.03) \times 10^{-1}$	
\\
\hline 

\multirow{4}{*}{Filaments}
&\multicolumn{1}{c|}{[9.5,10)}
& $(-0.30 \pm 0.16) \times 10^{-1}$	
& - 			
\\
&\multicolumn{1}{c|}{[10,10.5)}
& $(-0.37 \pm 0.05) \times 10^{-1}$	
& $(-0.29 \pm 0.03) \times 10^{-1}$	
\\
&\multicolumn{1}{c|}{[10.5,11.5]}
& - 	
& $(-0.30 \pm 0.02) \times 10^{-1}$	
\\
\noalign{\smallskip}
&\multicolumn{1}{c|}{$\Delta$}
& $(-0.26 \pm 0.06) \times 10^{-1}$	
& $(-0.21 \pm 0.02) \times 10^{-1}$				
\\
\hline 

\multirow{4}{*}{Walls}
&\multicolumn{1}{c|}{[9.5,10)}
& $(-0.34 \pm 0.19) \times 10^{-1}$	
& - 		 \\
&\multicolumn{1}{c|}{[10,10.5)}
& $(-0.31 \pm 0.07) \times 10^{-1}$	
& $(-0.33 \pm 0.05) \times 10^{-1}$			
\\
&\multicolumn{1}{c|}{[10.5,11.5]}
& - 	
& $(-0.32 \pm 0.02) \times 10^{-1}$			
\\
\noalign{\smallskip}
&\multicolumn{1}{c|}{$\Delta$}
& $(-0.21 \pm 0.09) \times 10^{-1}$	
& $(-0.25 \pm 0.02) \times 10^{-1}$	
\\

\hline
\end{tabular}
\end{table}

\begin{table}
	\centering
	\caption{Same as Table~\ref{tab:Gradients_sSFR} but for the gradients of the average stellar metallicities of sample centrals $\nabla _k ^{Z}$ and the gradients of the metallicity offset $\nabla _k ^{\Delta Z}$. The values are given in units of dex/dex. Grey numbers indicate values that are not significant at the $1\sigma$ level.}
	\label{tab:Gradients_Z}
	\begin{tabular}[!htb]{*{9}{c}}

&\multicolumn{1}{c|}{$M_\star$} & \multicolumn{1}{c}{field}
            & \multicolumn{1}{c}{groups}\\
                            
\hhline{====}

\multirow{4}{*}{Nodes}
&\multicolumn{1}{c|}{[9.5,10)}
& $(-0.53 \pm 0.19) \times 10^{-1}$	
& - 			 \\
&\multicolumn{1}{c|}{[10,10.5)}
& $(-0.16 \pm 0.1) \times 10^{-1}$	
& $(-0.27 \pm 0.06) \times 10^{-1}$	
\\
&\multicolumn{1}{c|}{[10.5,11.5]}
& - 	
& $(-0.18 \pm 0.02) \times 10^{-1}$				
\\
\noalign{\smallskip}
&\multicolumn{1}{c|}{$\Delta$}
& $(-0.12 \pm 0.08) \times 10^{-1}$	
& $(-0.14 \pm 0.03) \times 10^{-1}$			
\\
\hline 

\multirow{4}{*}{Filaments}
&\multicolumn{1}{c|}{[9.5,10)}
& \textcolor{gray}{$(-0.06 \pm 0.16) \times 10^{-1}$}	
& - 	
\\
&\multicolumn{1}{c|}{[10,10.5)}
& $(-0.17 \pm 0.05) \times 10^{-1}$	
& $(-0.21 \pm 0.04) \times 10^{-1}$		\\
&\multicolumn{1}{c|}{[10.5,11.5]}
& - 	
& $(-0.20 \pm 0.02) \times 10^{-1}$						
\\
\noalign{\smallskip}
&\multicolumn{1}{c|}{$\Delta$}
& \textcolor{gray}{$(-0.05 \pm 0.05) \times 10^{-1}$}	
& $(-0.13 \pm 0.01) \times 10^{-1}$	
\\
\hline 

\multirow{4}{*}{Walls}
&\multicolumn{1}{c|}{[9.5,10)}
& \textcolor{gray}{$(-0.04 \pm 0.18) \times 10^{-1}$}	
& - 		    	 \\
&\multicolumn{1}{c|}{[10,10.5)}
& $(-0.23 \pm 0.06) \times 10^{-1}$	
& $(-0.26 \pm 0.06) \times 10^{-1}$			\\
&\multicolumn{1}{c|}{[10.5,11.5]}
& - 	
& $(-0.18 \pm 0.02) \times 10^{-1}$			
\\
\noalign{\smallskip}
&\multicolumn{1}{c|}{$\Delta$}
& $(-0.15 \pm 0.09) \times 10^{-1}$	
& $(-0.14 \pm 0.02) \times 10^{-1}$				
\\

\hline
\end{tabular}
\end{table}

\begin{table}
	\centering
	\caption{Same as Table~\ref{tab:Gradients_sSFR} but for the gradients of the average stellar [$\alpha / \textrm{Fe}$] of sample centrals $\nabla _k ^{[\alpha / \textrm{Fe}]}$ and the gradients of the [$\alpha / \textrm{Fe}$] offsets $\nabla _k ^{\Delta [\alpha / \textrm{Fe}]}$. The values are in units of dex/dex. Grey numbers indicate values that are not significant at the $1\sigma$ level.}
	\label{tab:Gradients_afe}
	\begin{tabular}[!htb]{*{9}{c}}

&\multicolumn{1}{c|}{$M_\star$} & \multicolumn{1}{c}{field}
            & \multicolumn{1}{c}{groups}\\
                            
\hhline{====}

\multirow{4}{*}{Nodes}
&\multicolumn{1}{c|}{[9.5,10)}
& \textcolor{gray}{$(-0.15 \pm 0.16) \times 10^{-1}$}	
& - 	\\
&\multicolumn{1}{c|}{[10,10.5)}
& $(-0.19 \pm 0.08) \times 10^{-1}$	
& \textcolor{gray}{$(-0.02 \pm 0.07) \times 10^{-1}$}			
\\
&\multicolumn{1}{c|}{[10.5,11.5]}
& - 	
& $(-0.13 \pm 0.04) \times 10^{-1}$			
\\
\noalign{\smallskip}
&\multicolumn{1}{c|}{$\Delta$}
& $(-0.14 \pm 0.08) \times 10^{-1}$	
& $(-0.09 \pm 0.04) \times 10^{-1}$		
\\
\hline

\multirow{4}{*}{Filaments}
&\multicolumn{1}{c|}{[9.5,10)}
& \textcolor{gray}{$(-0.01 \pm 0.13) \times 10^{-1}$}	
& - 	 \\
&\multicolumn{1}{c|}{[10,10.5)}
& \textcolor{gray}{$(-0.01 \pm 0.05) \times 10^{-1}$}	
& $(-0.07 \pm 0.03) \times 10^{-1}$			\\
&\multicolumn{1}{c|}{[10.5,11.5]}
& - 	
& $(-0.21 \pm 0.02) \times 10^{-1}$		
\\
\noalign{\smallskip}
&\multicolumn{1}{c|}{$\Delta$}
& \textcolor{gray}{$(-0.01 \pm 0.06) \times 10^{-1}$}	
& $(-0.13 \pm 0.02) \times 10^{-1}$	

\\
\hline

\multirow{4}{*}{Walls}
&\multicolumn{1}{c|}{[9.5,10)}
& \textcolor{gray}{$(-0.05 \pm 0.13) \times 10^{-1}$}	
& - 		\\
&\multicolumn{1}{c|}{[10,10.5)}
& $(-0.16 \pm 0.06) \times 10^{-1}$	
& $(-0.14 \pm 0.06) \times 10^{-1}$		\\
&\multicolumn{1}{c|}{[10.5,11.5]}
& - 	
& $(-0.26 \pm 0.03) \times 10^{-1}$				
\\
\noalign{\smallskip}
&\multicolumn{1}{c|}{$\Delta$}
& $(-0.14 \pm 0.07) \times 10^{-1}$	
& $(-0.18 \pm 0.04) \times 10^{-1}$			
\\

\hline
\end{tabular}
\end{table}


\bsp	
\label{lastpage}
\end{document}